\documentclass[11pt]{article}
\usepackage[utf8]{inputenc}
\usepackage[T1]{fontenc}
\usepackage{epsfig}
\usepackage{color}
\usepackage{url}

\usepackage{breakurl}
\usepackage{booktabs, comment}       
\usepackage{amsfonts}       
\usepackage{graphicx,epsf,subcaption,mathrsfs}
\usepackage{amsmath,amssymb} %
\usepackage{enumitem}
\usepackage{mathrsfs}
\usepackage[title,titletoc,toc]{appendix} 
\usepackage[square,sort,comma,numbers]{natbib}
\usepackage{algorithm}
\usepackage{algorithmic}
\usepackage{setspace}
\usepackage{dsfont}
\usepackage[margin=1in,top=1in]{geometry}

\newtheorem{lemma}{\textbf{Lemma}}
\newtheorem{theorem}{\textbf{Theorem}}
\newtheorem{definition}{\textbf{Definition}}

\newtheorem{proposition}{\textbf{Proposition}}

\newcommand{\be}{\begin{eqnarray}}
	\newcommand{\ee}{\end{eqnarray}}
\newcommand{\ba}{\begin{eqnarray*}}
	\newcommand{\ea}{\end{eqnarray*}}
\newcommand{\bei}{\begin{itemize}}
	\newcommand{\beiftnt}{\begin{itemize}\footnotesize}
		\newcommand{\eei}{\end{itemize}}

	\def\mathbold{\boldsymbol}

	\def\bX{\mathbf{X}}
	\def\bW{\mathbf{W}}

	\newcommand {\bbeta} {\mbox{\boldmath $\beta$}}

	\def\bx{\mathbf{x}}
	\def\bz{\mathbf{z}}
	
	\def\bt{\mathbf{t}}
	
	\def\fm{\mathfrak{m}}

	\def\olL{{\overline L}}

	\def\bv{\mathbold{v}}

	\def\bZ{\mathbf{Z}}
	
	\def\bV{\mathbf{V}}
	
	\def\bY{\mathbf{Y}}

	\newtheorem{innercustomdef}{Definition}
\newenvironment{customdef}[1]
{\renewcommand\theinnercustomdef{#1}\innercustomdef}
{\endinnercustomdef}

\newtheorem{innercustomlem}{Lemma}
\newenvironment{customlem}[1]
{\renewcommand\theinnercustomlem{#1}\innercustomlem}
{\endinnercustomdef}

\begin{document}

\title{\vspace{-15mm}Distribution-Invariant Differential Privacy
}

\author{Xuan Bi$^1$ and Xiaotong Shen$^2$\thanks{$^1$Information and Decision Sciences, Carlson School of Management, University of Minnesota, Minneapolis, MN, email: xbi@umn.edu; $^2$School of Statistics, University of Minnesota, Minneapolis, MN, email: xshen@umn.edu. 
		}}

\date{}


\maketitle

\centerline{\bf Abstract}

{\footnotesize 
	Differential privacy is becoming one gold standard for protecting the privacy of publicly shared data. It has been widely used in social science, data science, public health, information technology, and the U.S. decennial census. Nevertheless, to guarantee differential privacy, existing methods may unavoidably alter the conclusion of original data analysis, as privatization often changes the sample distribution. This phenomenon is known as the trade-off between privacy protection and statistical accuracy. In this work, we {mitigate} this trade-off by developing a distribution-invariant privatization (DIP) method to reconcile both high statistical accuracy and strict differential privacy. As a result, any downstream statistical or machine learning task yields essentially the same conclusion as if one used the original data. Numerically, under the same strictness of privacy protection, DIP achieves superior statistical accuracy in a wide range of simulation studies and real-world benchmarks.
}

\vspace{.1in}

\noindent
{\footnotesize Key words: Privacy protection, distribution preservation, data sharing, data perturbation, randomized mechanism.\\
JEL classification code: C10}

\onehalfspacing


\section{INTRODUCTION}
Data privacy has become increasingly important in many fields in the big data era \citep{kearns2016private,cohen2020towards}, where a massive amount of sensitive information is digitally collected, stored, transferred, and analyzed. To protect data privacy, differential privacy \citep{dwork2006differential} has recently drawn great attention. It quantifies the notion of privacy for downstream machine learning tasks \citep{jordan2015machine} and protects even the most extreme observations. This quantification is useful for publicly released data such as {census} and survey data, and improves transparency and accessibility in artificial intelligence \citep{haibe2020transparency}. It has been adopted in biomedical research \citep{kaissis2020secure,jobin2021ai,hie2018realizing,han2020breaking}, epidemiology \citep{venkatramanan2021forecasting}, sociology \citep{santos2020differential}, and by many technology companies, such as Google \citep{erlingsson2014rappor}, Apple \citep{apple2017learning}, Microsoft \citep{ding2017collecting}, LinkedIn \citep{kenthapadi2018pripearl}, Amazon \citep{day2018protecting}, and Facebook \citep{nayak2020new}.
In 2020, differential privacy is, for the first time, used to protect the confidentiality
of individuals in the U.S. decennial census \citep{census2020disclosure}.

Scientists have developed various differential private methods. These methods 
not only protect data privacy but also promote data sharing. This allows privatized data from multiple
entities to be integrated into a single model to strengthen data analysis without leaking
critical information \citep{vadhan2017complexity}. This aspect brings huge economic and societal benefits in the big data era.
However, one major issue is that data privatization may alter the analysis result of the original data, and it is believed that there is a trade-off between statistical accuracy and differential privacy \citep{goroff2015balancing,santos2020differential,gong2020congenial,bowen2020comparative}. In other words, one needs to sacrifice the accuracy of a downstream analysis for privacy protection. On the contrary, we show that one can reconcile \emph{both} accuracy and privacy, which we achieve by preserving the original data's distribution. 
We develop a distribution-invariant privatization method (DIP) that achieves private data release. 
It transforms and perturbs the data and employs a suitable transformation to recover the original distribution. As a result, DIP maintains statistical accuracy while being differentially private at any desired level of protection.

There is a large body of literature on differential privacy.
Two major directions emerge from computer science and statistics. The
first achieves privacy protection by either a privatization mechanism
or a privatized sampling method,
including the Laplace mechanism \citep{dwork2006calibrating,dwork2014algorithmic}, the exponential
mechanism \citep{mcsherry2007mechanism}, the minimax optimal procedures \citep{duchi2018minimax}, among others. 
The second achieves differential privacy via privatization
for a category of models or algorithms, such as deep learning \citep{abadi2016deep}, boosting \citep{dwork2010boosting}, stochastic gradient descent \citep{agarwal2018cpsgd}, risk minimization \citep{chaudhuri2011differentially}, random graphs \citep{karwa2016inference}, function estimation \citep{hall2013differential}, parametric estimation \citep{avella2019privacy}, regression diagnostics \citep{chen2016differentially}, and top-$k$ selection \citep{durfee2019practical}.
Moreover, a general statistical framework for differential privacy is introduced \citep{wasserman2010statistical}, and theoretical properties of various privatization mechanisms are investigated \citep{kairouz2017composition, vadhan2017complexity}. Other types of differential privacy as well as the associated
statistical properties are also discussed, including relaxed or approximate differential privacy \citep{dwork2006our,abadi2016deep,agarwal2018cpsgd,cai2019cost}, local differential privacy \citep{evfimievski2003limiting,kasiviswanathan2011can,ding2017collecting,rohde2018geometrizing}, random differential privacy \citep{hall2012random}, Renyi differential privacy \citep{mironov2017renyi}, and Gaussian differential privacy \citep{dong2019gaussian}.

One main challenge is that existing privatization mechanisms protect data privacy at the expense of altering a sample's distribution. 
For example, count data may become negative values after adding a noise, {and non-trivial or data-dependent post-processing may be needed}. 
As a result, analysis of such privatized data may conclude dramatically different from the analysis of the original data. From a machine learning perspective, developing a distribution-invariant privatization mechanism becomes crucial to maintain statistical accuracy.

This article proposes a DIP mechanism to address the above challenge for essentially \emph{all} types of data, such as those following continuous, discrete, mixed, categorical, empirical, or multivariate distributions.
It satisfies differential privacy while approximately preserving the
original distribution when it is unknown through a reference distribution, 
for example, the empirical distribution on a hold-out sample and exactly preserving the original distribution when it is known, c.f., Theorems \ref{dp_multivar} and \ref{consistency_multivar}. It is scalable to massive or high-dimensional data, c.f., Proposition \ref{prop:multi_complexity}.
Consequently, any downstream statistical analysis or machine learning tasks will lead to nearly the same conclusion as if the original data were used, which is a unique aspect that existing methods do not share.
It allows distributions with unbounded support, which would be otherwise rather difficult if not impossible
\citep{wasserman2010statistical}.
Moreover, DIP approximately maintains statistical accuracy even with strict privacy protection in that it does not suffer from the trade-off between accuracy and privacy 
strictness, as illustrated in our simulation studies in Section 2. 
These characteristics enable us to perform data analysis without 
sacrificing statistical accuracy,
as in regression, classification, graphical models, clustering, among other statistical and machine learning tasks.

\section{RESULTS}

\subsection{Overview of the Proposed Method} \label{sec:method}

This subsection presents the main ideas of the proposed DIP method. Differential privacy protects 
publicly shared information about a dataset by describing its patterns while guarding against disclosing the data information. The reader may consult Definition \ref{DP} for a rigorous mathematical definition. In what is to follow,
we discuss the privatization of univariate and multivariate samples separately. 

DIP's privatization process consists of three steps. First, DIP splits the
original sample randomly into two independent subsamples, hold-out and to-be-privatized samples, both are fixed after the split.  Second, it estimates an unknown data distribution by, say,  the empirical distribution on the hold-out sample, which is referred to as a reference distribution. Third, we privatize the to-be-privatized sample through data perturbation, which (i) satisfies the requirement of differential privacy, and (ii) preserves the reference distribution approximating the original distribution. As a result, DIP is differential private on the to-be-privatized sample while retaining the original distribution asymptotically, c.f., Theorem \ref{consistency_multivar}.

For univariate data, the privatization process of a to-be-privatized sample is described in Figure \ref{flowchart} and Section \ref{sec4}. First, we apply the probability-integral transformation to each data point, or its smooth version of the reference distribution, to yield a uniform distribution. Second, we add a random Laplace noise to perturb and mask the data, where the Laplace noise entails differential privacy \citep{dwork2006differential}, as opposed to, say, the Gaussian noise. Third, we design a new function transforming the obfuscated data to follow the reference distribution approximating the original
data distribution.

For multivariate data, we privatize each variable sequentially, using the univariate method above. To preserve the distribution, we propose to apply the probability chain rule \citep{schum2001evidential}, in place of privatizing each variable independently. In other words, we privatize the first variable, and then the second variable using its conditional distribution given the privatized value of the first variable, and so forth;
c.f.,  {Section \ref{sec4}}.


Finally, we develop a fast and scalable algorithm, c.f., Algorithm \ref{alg:multivar} and Proposition \ref{prop:multi_complexity} in Section \ref{sec4}, for practical use, and prove that DIP preserves the data distribution 
approximately while being differentially private for essentially all 
types of data, c.f., Theorem \ref{consistency_multivar}.


%

\subsection{Simulation Studies}

This subsection performs simulations to investigate the operating characteristics of DIP and compare it with some top competitors, including
the Laplace randomized mechanism (LRM) \citep{dwork2006calibrating}, the minimax optimal procedure mechanism (OPM) \citep{duchi2018minimax},
and the exponential mechanism (EXM) \citep{mcsherry2007mechanism}. 
For a fair comparison, all competing methods, including DIP, use the same original dataset and the same privacy factor, while assuming the data distribution is unknown. Then the statistical accuracy of a downstream method on privatized data is evaluated. It is worth mentioning that DIP requires sample-splitting, which is at a disadvantage for downstream analysis due to the reduced size of the released sample.
Despite the disadvantage, DIP achieves superior statistical accuracy in two simulations and on three real-world benchmarks.

%
%
%
%
%

\subsubsection{Network structure reconstruction} \label{sec:reg}

We first consider the private reconstruction of a network's structure. We generate a random sample from a multivariate normal distribution, in which its precision matrix, or the inverse of the covariance matrix, encodes an undirected graph 
for the network. Then we estimate the precision matrix based on the privatized 
random sample.

We examine two types of graph networks, including the chain and exponential
decay networks. For the chain network, the precision matrix is a sparse tridiagonal matrix 
corresponding to the first-order autoregressive structure. 
For the exponential decay network, elements of 
the precision matrix are exponentially decayed, and the network is not sparse.
We perform 1000 simulations in a high-dimensional setting where
the sample size $N=200$ is smaller than the number of variables {$p=250$}.
The privacy factor is from 1 to 3. 
After privatization, we estimate the precision matrix via
graphical Lasso \citep{friedman2008sparse}. The estimated precision matrix 
is evaluated by the entropy loss \citep{lin1985monte} by comparing it with the true precision matrix.
For DIP, we apply \eqref{multi} with a random split of the original
sample where a ratio of $15\%, 25\%, 35\%$ is used as the hold-out sample.
For LRM and OPM, we follow \citep{dwork2006calibrating,duchi2018minimax}
to privatize the entire sample.
 

As suggested by Table \ref{sim2:graphical}, 
DIP performs well. DIP is insensitive to the value of the privacy factor, which agrees with the theoretical result in Theorem \ref{consistency_multivar}. 
Moreover, 
neither LRM nor OPM applies because of the requirement of bounded support, as evident by an infinite value of the entropy loss caused by the estimate of the precision matrix being 0. In summary, DIP's distribution preservation property becomes more critical to explore multivariate structures, which explains sizable differences between DIP and its competitors (LRM and OPM) in simulations.

%
%
%

\subsubsection{Linear regression}

Consider a linear model in which the sample size is $N=200$ or $2000$ and the design matrix consists of 6 or 30 variables.
The regression coefficient is set as a vector of 1's. Each 1/3 of the design matrix's columns follow independently $Normal(0,10^2)$, $Poisson(5)$, and $Bernoulli(0.5)$, respectively. The privacy factor is $1,2,3$, or $4$.
In this example, we estimate the regression coefficient
vector-based on the privatized release data
and measure the estimation accuracy by the Euclidean distance between the estimated and true parameter vectors.

For DIP, we split the data into hold-out and to-be-privatized samples with a splitting
ratio of 15\%, 25\%, and 35\%. Then we apply Algorithm \ref{alg:multivar} to privatize
variables of the to-be-privatized sample in a random order to examine DIP's invariant property
of the sequential order in Theorem \ref{consistency_multivar}.
For LRM and OPM, we adopt the univariate case as in the graphical model case.
Note that LRM and OPM utilize the additional information -- independence among columns of the design matrix, whereas
DIP does not.
The {simulation} replicates 1000 times.

As indicated in Table \ref{sim2:linear}, DIP yields the lowest estimation
error across all situations with a substantial amount of improvement over LRM 
and OPM, ranging from 559\% to 1071503\%, despite that DIP's to-be-privatized
sample size is smaller than the size of the entire sample. Importantly, DIP {mitigates} the trade-off between differential privacy and statistical accuracy, as an increased level of the privacy factor has little impact on the estimation accuracy.
Meanwhile, only a small difference 
is seen between DIP and the oracle non-private counterpart. This
is a small price to be paid for estimating an unknown distribution.
The small standard error also suggests that the multivariate DIP is invariant against the random order of variable privatization,
which agrees with the invariant property for the sequential order in Theorem \ref{consistency_multivar}. 

\subsection{Real Data Analysis} \label{sec:real}

Next, we analyze three sensitive benchmark datasets to understand the practical implications of distribution-invariant privatization. 

\subsubsection{The University of California salary data}

The first study concerns the University of California system salary data collected in 2010 \citep{uc2014annual}. 
The dataset contains annual salaries of 252,540 employees, 
including faculty, researchers, and staff.
The average salary of all employees is \$39,531.49 with a 
standard deviation of \$53,253.93. The data distribution is highly right-skewed, with
the 90\% quantile being \$95,968.12 and the maximum exceeding two million dollars.

For this study, we estimate the differentially private mean salary.
One important aspect is contrasting the privatized mean with the original mean
\$39,531.49 to understand the impact of privatization on statistical accuracy
of estimation. The relative mean difference (i.e., the difference between the private mean and the original mean divided by the original mean) is evaluated. Three privatization mechanisms are compared, including
DIP, LRM, and OPM. For DIP, we apply Algorithm \ref{alg:multivar}. For LRM, random Laplace noise is added to the original data before calculating the private mean. For OPM, we follow the private mean estimation function described in Section 3.2.1 of \cite{duchi2018minimax} to optimize its performance. The above process, including privatization, is repeated 1000 times.


\subsubsection{Portuguese bank marketing campaign data}

 The second study focuses on a set of marketing campaign data collected from a Portuguese retail bank from 2008 to 2013 \citep{moro2014data}.
This campaign intended to sell long-term deposits to potential clients through phone conversations. During a phone call, an agent collected a client's sensitive personal information, including age, employment status, marital status, education level, whether the client has any housing or personal loan, and if the client is in a default status. In addition, the agent collected the client's device type, past contact histories regarding this campaign, and if the client is interested in subscribing to a term deposit (yes/no). The dataset contains a total of 30,488 respondents whose data are complete.

Our goal is to conduct private logistic regression and examine the statistical accuracy change after privatization.
For DIP, Algorithm \ref{alg:multivar} is applied. For LRM, Laplace noise is added to each variable independently. For OPM, we conduct private logistic regression following the private estimation of generalized linear models in Section 5.2.1 of \cite{duchi2018minimax}. The Kullback-Leibler divergence is measured to evaluate the discrepancy between the private and the original logistic regression results. The privatization process repeats 1000 times.


\subsubsection{MovieLens data}

The third study considers the privacy protection of movie ratings for a private recommender system. Many online platforms disclose de-identified user data for research or commercial purposes. In many situations,
however, simply removing user identities is not
adequate. For example, as suggested in \citep{narayanan2008robust},
political preferences and other sensitive information
can still be uncovered based on de-identified movie ratings.

To overcome this difficulty, we consider the MovieLens 25M dataset, which is the most recent dataset collected by GroupLens Research \cite{harper2015movielens} between 1995 and 2019. It contains 25,000,095 movie ratings, collected from 162,541 users over 59,047
movies. In other words, each user rated 154 out of the 59,047 movies on average, leaving their preference towards the vast majority of movies unknown. Movie ratings are values in $\{0.5,1,1.5,\ldots,5\}$. A typical recommender system intends to predict a user's rating on movies that they have not watched yet, and then recommends movies to each user based on high predicted ratings. The focus here 
is a prediction as opposed to inference in the previous two studies. 

To investigate the effect of data privatization on the prediction accuracy, we randomly split the movie ratings into a 75\% training set and a 25\% test set, and privatize the training set. For DIP, we apply Algorithm \ref{alg:multivar}. For LPM, Laplace noises are added to the raw ratings, and the noised ratings are rounded to the nearest number in $\{0.5,1,1.5,\ldots,5\}$. For OPM, we follow Section 4.2.3 of \cite{duchi2018minimax}. For EXM, the sampling scheme described in Appendix S2 (in the subsection about the MovieLens data) is applied. 
Then we train a matrix factorization model \citep{funk2006netflix} based 
on privatized ratings, which is a prototype collaborative filtering method 
for movie recommendation. The evaluation metric is the root mean square error
on the non-private test set, which is averaged over 50 random partitions of training and test sets. 

\subsubsection{Analysis results}
In all three benchmark examples, the privacy factor is 1. 
For DIP, we hold out $25\%$ of the sample for estimating the unknown data distribution.
As shown in Table \ref{real_logistic}, DIP delivers desirable statistical
accuracy across the three different statistical and machine learning tasks,
demonstrating the benefits of distribution-preservation.

For the mean salary estimation in the first example, the discrepancy between the original mean and the private mean of DIP is minimal. Specifically, while holding the same level of strictness on privacy protection, DIP only entails an error of about 0.4\% of the original sample mean, and the error is 13.08\% and 4.69\% for LRM and OPM, respectively.
And the amount of improvement of DIP over LRM and OPM is 3170.0\% and 1072.5\%. In other words, DIP achieves the highest accuracy while guaranteeing differential privacy.

For logistic regression in the second example, DIP entails a small error, suggesting that any conclusion based on DIP's private regression would remain nearly the same as the one based on the original data. In terms of statistical accuracy, DIP delivers a substantial improvement of 561.7\% over LRM,  whereas OPM fails to produce any meaningful results due to an inaccurately estimated probability of 1 or 0 across all settings.

For personalized recommendations in the third example, DIP performs the best and yields a significant improvement of 81.6\%, 153.4\%, and 7.8\% over LRM, OPM, and EXM, respectively. This result implies that, while protecting data privacy, preserving the original data distribution also ensures high prediction accuracy.
Additional supporting numerical evidence is given in Appendix S2.

%
%
%

\section{DISCUSSION}

Differential privacy has become a standard of data privacy protection,   
as a large amount of sensitive information is collected and stored in a digital form. This paper proposes a novel privatization method, DIP, which preserves the original data's distribution while satisfying differential privacy. DIP {mitigates} the trade-off between data privacy and statistical accuracy. Consequently, any downstream privatized statistical analysis or machine learning task leads to the same conclusion as if the original data were used, which is a unique aspect that all existing mechanisms may not enjoy.
Second, DIP is differentially private even if underlying data have
unbounded support or unknown distributions.
Our extensive numerical studies demonstrate the utility of DIP against top competitors across many distinct application scenarios.
On the other hand, DIP requires a hold-out sample to estimate the data's distribution. This may
reduce the released sample size due to sample splitting.

The proposed methodology also opens up several future fronts.
One is the generalization to
local differential privacy, which provides further privacy protection for data in a local device or server. Some discussions of DIP on local differential privacy are provided in Appendix S3. Another direction is the extension to independent but not identically distributed data, in which DIP can still 
ensure differential privacy while preserving the distribution, but requires repeated measurements for each individual. See Appendix S4 for more details.


\section{METHODS}
\label{sec4}

\subsection{Differential Privacy}

%


Differential privacy ensures that the substitution of any single observation in a dataset would have a small impact on the publicly shared information, which can be measured by $\varepsilon$-differential privacy.
Suppose $\bZ$ is a random sample from a cumulative distribution function (cdf) $F$, and $\mathfrak{m}$ is a \emph{ privatization mechanism} which privatizes $\bZ$ into $\tilde{\bZ}$ for public release. Let $\bz$ and $\bz'$ be two \emph{adjacent} realizations of $\bZ$, which differ in only one observation.
Then $\varepsilon$-differential privacy requires that the ratio of the probability of any privatized event given one sample to the probability of it given the other sample is upper bounded by $e^{\varepsilon}$, that is:
\begin{definition} \citep{dwork2006calibrating, dwork2006differential}
	\label{DP}
	A privatization mechanism $\fm(\cdot)$ satisfies $\varepsilon$-differential privacy if
	\ba
	\sup_{\bz,\bz'}\sup_B\frac{P\big(\fm(\bZ) \in B|\bZ=\bz\big)}{P\big(\fm(\bZ)\in B|\bZ=\bz'\big)} \le e^{\varepsilon},
	\ea
	where $B$ is a measurable set and
	$\varepsilon \ge 0$ is a privacy factor that is usually small. For convenience,
	the ratio is defined as 1 when the numerator and denominator are 0.
\end{definition}

Definition \ref{DP} requires that the ratio of conditional probabilities
of any privatized event (i.e., the set $B$) given two \emph{adjacent} data realizations
is no greater than $e^{\varepsilon}$. Here $\varepsilon$ is called
a \emph{privacy factor}, which characterizes the budget of privacy protection. For example, a small value of $\varepsilon$ renders a strict privacy protection policy. 

\subsection{Implications of Differential Privacy}

Differential privacy protects against any data identification or revealing data values of
the original sample, by adversary attack. However, the disclosure odds can increase as
an adversary attack repeats multiple times. Subsequently, 
to understand the level of protection of differential privacy, as specified by the privacy factor $\varepsilon$ in Definition \ref{DP}, we generalize Theorem 2.4 of \cite{wasserman2010statistical} to 
a repeated adversary attack.

Consider a repeated adversary attack, where an adversary makes $M$ queries to obtain $M$ independent $\varepsilon$-differentially private samples, intending to reveal the values of an original 
sample $\bZ=(Z_1,\cdots,Z_n)$. Lemma \ref{rep_queries} details the level of protection by $\varepsilon$-differential privacy 
against data identification in a repeated adversary attack. 

\begin{lemma} \label{rep_queries} 
	Suppose $\tilde{\bZ}^{(1)},\ldots,\tilde{\bZ}^{(M)}$ are $M$ independent yet $\varepsilon$-differentially private copies of $\bZ$. 
	Then any hypothesis test to identify the value of the $i_0$th observation $Z_{i_0}$, namely $H_0: Z_{i_0}=\mu_0$ versus $H_1: Z_{i_0} =\mu_1$,
	based on $\tilde{\bZ}^{(1)}, \ldots, \tilde{\bZ}^{(M)}$ has statistical power no greater than
	$\gamma e^{M\varepsilon}$ with any $\mu_1 \neq \mu_0$ and any $i_0=1,\ldots,n$, given a significance level $\gamma>0$.
\end{lemma}

In particular, Lemma \ref{rep_queries} states that it is impossible to reject a null hypothesis $H_0$ that an observation equals a specific value
$\mu_0$ in any sample because of a small power $\gamma e^{M\varepsilon}$ for sufficiently small
$\varepsilon$, especially so in a one-time data-release scenario with $M=1$. 
However, the information leak could occur when $M$ increases, 
while holding $\varepsilon$ fixed in that $H_0$ is eventually rejected as a result of increased power.

In summary, $\varepsilon$-differential privacy only protects against data identification
when $M$ is limited and an adversary attack will eventually break the defense of 
$\varepsilon$-differential privacy as the number of attacks $M$ increases. In practice,
$M$ needs to be limited for $\varepsilon$-differential privacy depending on
the protection factor $\varepsilon$.

Next, we show, in Lemma \ref{direct_ptb}, that any linear perturbation or 
transformation does not entail $\varepsilon$-differential privacy for
the original data with unbounded support. This means that we must seek nonlinear transformations
beyond the linear domain to protect data with unbounded support, which 
motivates the proposed method \eqref{ptb_c}. 

\begin{lemma} \label{direct_ptb} For any multivariate random sample $(\bZ_1,\ldots, \bZ_n)$,  
	any linear privatization mechanism $\fm(\cdot)$: $\fm(\bZ_i)=\bbeta_0+\bbeta_1 \bZ_i 
	+\mathbf{e}_i$; $i=1,\ldots,n$, is not $\varepsilon$-differentially private when $\bZ_i$ has unbounded support, 
	where $\bbeta_0$ and $\bbeta_1$ ($\bbeta_1 \ne 0$) are any fixed coefficients, and $\mathbf{e}_i$ is any
	random noise vector. 
\end{lemma}

\subsection{Theoretical Justification} \label{sec:true}

This subsection discusses DIP in the context of a known data distribution 
for motivation, which paves up the way for the theoretical 
justification of the distribution preservation property. Then, in
a subsequent subsection, we further expand the method to an unknown 
data distribution by estimating it based on a hold-out sample.

\emph{Univariate continuous distributions.} 
Suppose $(Z_1,\ldots,Z_n)$ is a random sample of a given cumulative distribution function $F$. We begin with our discussion with a known continuous $F$.

Our privatization proceeds by transforming
each $Z_i$ through the following steps, as displayed in  Figure \ref{flowchart}; $i=1,\ldots,n$. 
First, we apply $F$ to $Z_i$ to yield a uniformly distributed variable $F(Z_i)$. Second, we add an independent 
noise $e_i$ to perturb $F(Z_i)$ for privacy protection, where $e_i$ is randomly sampled
from a Laplace distribution $Laplace(0,1/\varepsilon)$.
Here the scale parameter $\varepsilon$ ensures that 
our privatization satisfies $\varepsilon$-differential privacy.
Finally, we apply a nonlinear transformation $H$ 
to produce a privatized sample that follows the original distribution $F$. 

The specific form of $H$ depends on the data type. 
For a univariate continuous sample, $H$ consists of two parts. The cumulative distribution function $G$ is applied to $F(Z_i)+e_i$ to produce a uniform variable, where $G$'s explicit expression is given in Appendix S1.1. Then, the inverse cumulative distribution function $F^{-1}$ is applied to obtain the original distribution. In summary,
we generate a DIP's privatized sample $(\tilde{Z}_1,\ldots,\tilde{Z}_n)$ via a formula:  
\be \label{ptb_c}
\tilde{Z}_i = H\big(F(Z_i)+e_i\big), \quad H(\cdot)=F^{-1}\circ G(\cdot),
\ee
where $\circ$ denotes function composition.


\emph{Univariate discrete distributions.} For a discrete or mixed variable, 
that is, a variable with both continuous and discrete components, \eqref{ptb_c} still applies if we replace $F$ by a smoothed $F$, a process we call continualization, as illustrated in Figure \ref{pdf_continualization}.
Specifically, we subtract a uniformly distributed random variable $U_i$ from $Z_i$, such that the distribution of the modified $Z_i$ follows the smoothed $F$, followed by privatization in \eqref{ptb_c}. Then, we apply a ceiling function such that the privatized sample follows the original distribution $F$, c.f.,
the Appendices S1.3 and S1.4.

\emph{Multivariate distributions.} Suppose $(\bZ_1,\ldots,\bZ_n)$ is a random sample of a multivariate distribution $F$, where $\bZ_i=(Z_{i1},\ldots,Z_{ip})$ is a $p$-dimensional vector.
We privatize $\bZ_i$ by applying \eqref{ptb_c}, directly or after continualization, to each variable sequentially via the probability chain rule. In particular, we privatize its first component $Z_{i1}$ 
to yield its privatized value $\tilde{Z}_{i1}$. Then privatize $Z_{i2}$ given $\tilde{Z}_{i1}$
using the conditional cumulative distribution $F(Z_{i2}|Z_{i1})$ and the inverse function $F^{-1}(Z_{i2}|\tilde{Z}_{i1})$ to replace $F$ and $F^{-1}$ in \eqref{ptb_c}, respectively, to 
yield $\tilde{Z}_{i2}$, and so forth. This leads to our
general formula for DIP:
\be
\label{multi}
\tilde{Z}_{i1}=\fm_{1}(Z_{i1}), \quad \tilde{Z}_{il} = \fm_{l}(Z_{il}|\tilde{Z}_{i1},\ldots,\tilde{Z}_{i,l-1});
\quad l=2,\ldots,p,
\ee
where $\fm_{l}$ denotes \eqref{ptb_c} for the $l$th variable, with $F$ in \eqref{ptb_c} (and the corresponding $F^{-1}$) replaced by the marginal distribution of $Z_{i1}$ for $l=1$, and the conditional distribution of $Z_{il}$ given $\tilde{Z}_{i1},\ldots,\tilde{Z}_{i,l-1}$ for $l=2,\ldots,p$. 

The properties for both univariate and multivariate privatizations are summarized in Theorem \ref{dp_multivar} below.

\begin{theorem} \label{dp_multivar} 
DIP in \eqref{multi} is $\varepsilon$-differentially private, and the privatized sample follows the original distribution $F$ when $F$ is known.
\end{theorem}


{It is important to note that DIP in \eqref{ptb_c} and \eqref{multi} differs substantially from a sampling method. A sampling method generates synthetic data from $F$, which does not require raw data $\bZ$ when $F$ is known. On the other hand, DIP preserves the data identifier $i$ of $\bZ_i$, $i=1,\ldots,n$. 
This property is useful for data integration or personalization. For example, when collecting an additional column of data from the same individual $i$, for example,
user $i$'s ratings on a new movie in the context of Section 2.3.3, we may integrate them with $\tilde \bZ_i$. In contrast,  a sampling method does not retain any data identifier.

\subsection{The Proposed Method} \label{sec:empirical}

In practice, $F$ is usually unknown and has to be replaced by an estimate $\hat{F}$, for example, the empirical cumulative distribution function. However, $\hat{F}$ needs to be independent of the privatized data to satisfy differential privacy. {Towards this end, we either (i) construct $\hat{F}$ based on a random subsample of $\bZ$, called a \emph{hold-out sample}. The hold-out sample is neither privatized nor released while the remaining sample is \emph{to-be-privatized}, c.f., Appendix S1.2; or (ii) use a public dataset that follows the same distribution $F$ as the hold-out sample, while treating the entire $\bZ$ as the to-be-privatized sample. For example, the American Community Survey data are public and the U.S. census data are private, both coming from the same population. The former can serve as a hold-out sample for the latter.} Here both the hold-out and the to-be-privatized samples are fixed once assigned. We then apply \eqref{multi} with
$F$ replaced by $\hat{F}$ to the to-be-privatized sample, which asymptotically preserves the original distribution when $\hat{F}$ is a consistent estimate of $F$, for example, {the smoothed empirical cumulative
distribution function}, c.f., Theorem \ref{consistency_multivar}. See Appendices S1.5 and S1.6 for the complete technical details of constructing the DIP mechanism for univariate and multivariate variables with $\hat{F}$, respectively.

Algorithm \ref{alg:multivar} summarizes the entire privatization process of DIP.
Let $N=n+m$ be the total sample size, where $m$ is the hold-out sample size. Proposition \ref{prop:multi_complexity} concerns the computational efficiency and scalability of Algorithm \ref{alg:multivar}.

\begin{algorithm}
	\caption{Distribution-invariant privatization}
	\label{alg:multivar}
	\begin{algorithmic}
		\STATE{{\bf Input}: A to-be-privatized sample $(\bZ_1,\ldots,\bZ_n)$, 
a hold-out sample, dimension $p$, the privacy factor $\varepsilon$.}
		\FOR{$ i= 1,\ldots,n$}
		\FOR{$ l= 1,\ldots,p$}
		\STATE{Continualize $Z_{il}$ if it is not continuous.}
		\STATE{Privatize $Z_{il}$ into $\tilde{Z}_{il}$ following \eqref{multi} with a privacy factor $\varepsilon/p$ and with $\hat{F}$ estimated by the hold-out sample.}	
		\ENDFOR
		\FOR{$ l= 1,\ldots,p$}
		\STATE{Apply a proper ceiling function to $\tilde{Z}_{il}$ if $Z_{il}$ is not continuous.}
		\ENDFOR
		\ENDFOR
		\STATE{{\bf Output}: A privatized sample $(\tilde{\bZ}_1,\ldots,\tilde{\bZ}_n)$.}
	\end{algorithmic}
\end{algorithm}


\begin{proposition} \label{prop:multi_complexity}
	The computational complexity of Algorithm \ref{alg:multivar} is $O(p N \log N)$.
\end{proposition}

Theorem \ref{consistency_multivar} establishes DIP's $\varepsilon$-differential privacy 
and its asymptotic distribution preservation as the size of the hold-out sample 
tends to infinity.

\begin{theorem} \label{consistency_multivar}
	DIP in Algorithm \ref{alg:multivar} is $\varepsilon$-differentially private, and the privatized sample follows the original distribution $F$ 
asymptotically {as $m \rightarrow \infty$} when $F$ is unknown.
\end{theorem}

%
%

To guarantee $\varepsilon$-differential privacy, 
we apply the sequential composition \citep{dwork2006calibrating,kairouz2017composition} and require each $\fm_{l}$ in \eqref{multi} to be 
$\varepsilon/p$-differentially private, which is enforced by
sampling $e_{il}$ from a Laplace distribution $Laplace(0,p/\varepsilon)$.
A small loss of statistical accuracy may incur depending on the estimation precision of $\hat{F}$ for $F$ in a finite-sample situation. Whereas a large $m$ provides a 
refined estimate of $F$, a large $n$ renders a large to-be-privatized sample for downstream statistical analysis.  In general, one may choose a reasonable $m$ to retain statistical accuracy. However, when $N=n+m$ is fixed,  one needs to consider if the released sample size is adequate.

As a technical note, the DIP result 
continues to hold for any consistent estimator of $F$, which is useful when additional information is available. 
For example, if $Z_i$'s follow a normal distribution $N(\mu,1)$ with an unknown
mean value $\mu$, then $\hat{F}$ can be chosen as the continuous cdf of a normal distribution with an estimated mean $\hat{\mu}$ and variance 1.

It is worth noting that the privacy of raw data in the hold-out sample is also protected when a public dataset is unavailable for estimating $F$. First, any transmission, alteration, querying, or release of any raw data in the hold-out sample is not permissible. 
Second, only a continuous version of the estimated $F$ is constructed, which 
guarantees that $\hat F$ does not contain any probability mass of the raw data. This is achieved through adding noise to the raw data in the hold-out sample. See Appendices S1.5 and S1.6 for more details. Therefore, the privacy protection of a hold-out sample remains a high standard.\footnote{Differential privacy as in Definition \ref{DP}
is not well defined on the hold-out sample due to no adjacent realizations being permissible.} As shown in  Lemma \ref{holdout}, the probability of identifying a continualized value of a hold-out sample from the privatized data is zero. In other words, there is
no privacy leakage of the hold-out sample. 

\begin{lemma} \label{holdout} For any individual $j$ in the hold-out sample, that is, $j=n+1,\ldots,n+m$, let $\bV_j$ be the continualized version of $\bZ_j$. Then $P(\bV_j=\bv|\tilde{\bZ}_1=\tilde{\bz}_1,\ldots,\tilde{\bZ}_n=\tilde{\bz}_n)=0$ for any $\bv,\tilde{\bz}_1,\ldots,\tilde{\bz}_n \in \mathbb{R}^p$.
\end{lemma}

The notion of ``hold-out'' is an analogy to that of retaining sensitive data confidential, as required by the National Institutes of Health (NIH) to perform medical research. In particular, NIH's policy states that personally identifiable, sensitive, and confidential information should ``not be housed on portable electronic devices'', and researchers should ``limit access to personally identifiable information through proper access controls.''\footnote{The original policy is available at \url{https://grants.nih.gov/grants/policy/nihgps/html5/section_2/2.3.12_protecting_sensitive_data_and_information_used_in_research.htm}}
Accordingly, the privacy of the hold-out sample here is protected in the same fashion -- ideally stored in a room  without Internet and prevented from unauthorized access, in addition to noise injection, continualization, and the requirement of not being transmitted, altered, queried, or released.


\section*{Acknowledgements}

The authors thank the editor and three reviewers for insightful comments and suggestions, which improve the article significantly. This research is supported in part by NSF grant DMS-1952539, and NIH grants R01AG069895, R01AG065636, 1R01GM126002, R01HL105397, 
R01AG074858, and U01AG073079.

\section*{Figures and Tables}

\begin{figure}[H]
	\centering
	\includegraphics[width=1\textwidth]{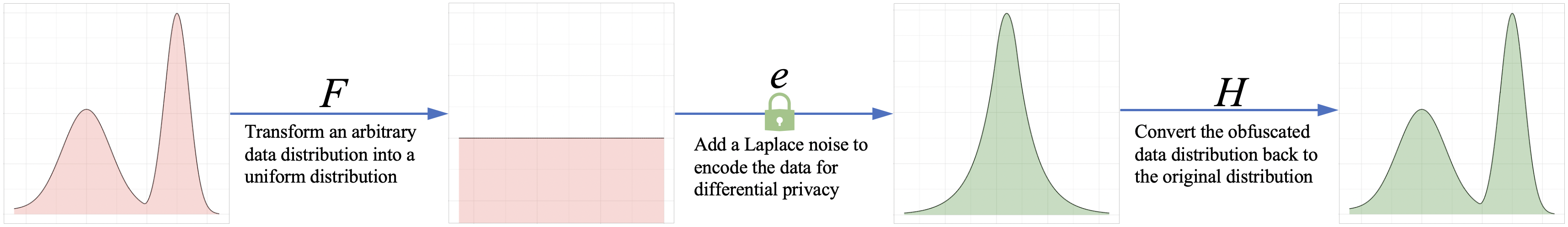}
	\caption{\textbf{A flowchart visualizing the proposed DIP method.} The left arrow: Data from an arbitrary distribution is transformed into a sample of the uniform distribution via the probability integral transformation function $F$. The middle arrow: A noise following a centered Laplace distribution is added to the transformed data to ensure differential privacy. The obfuscated data now follow a convoluted distribution, whose explicit form is provided in the Appendix. The right arrow: A carefully designed transformation function $H$ is applied to the obfuscated data to obtain a privatized sample, which follows the original distribution but does not necessarily contain the original data values. Non-private and private distributions are filled in red and green, respectively.}
	\label{flowchart}
\end{figure}

\begin{figure}[H]
	\centering
	
\makebox[\linewidth][c]{
	\vspace{-.03in}
	\begin{subfigure}{.33\textwidth}\hspace{0.8in}
		\centering
		\includegraphics[width=\textwidth]{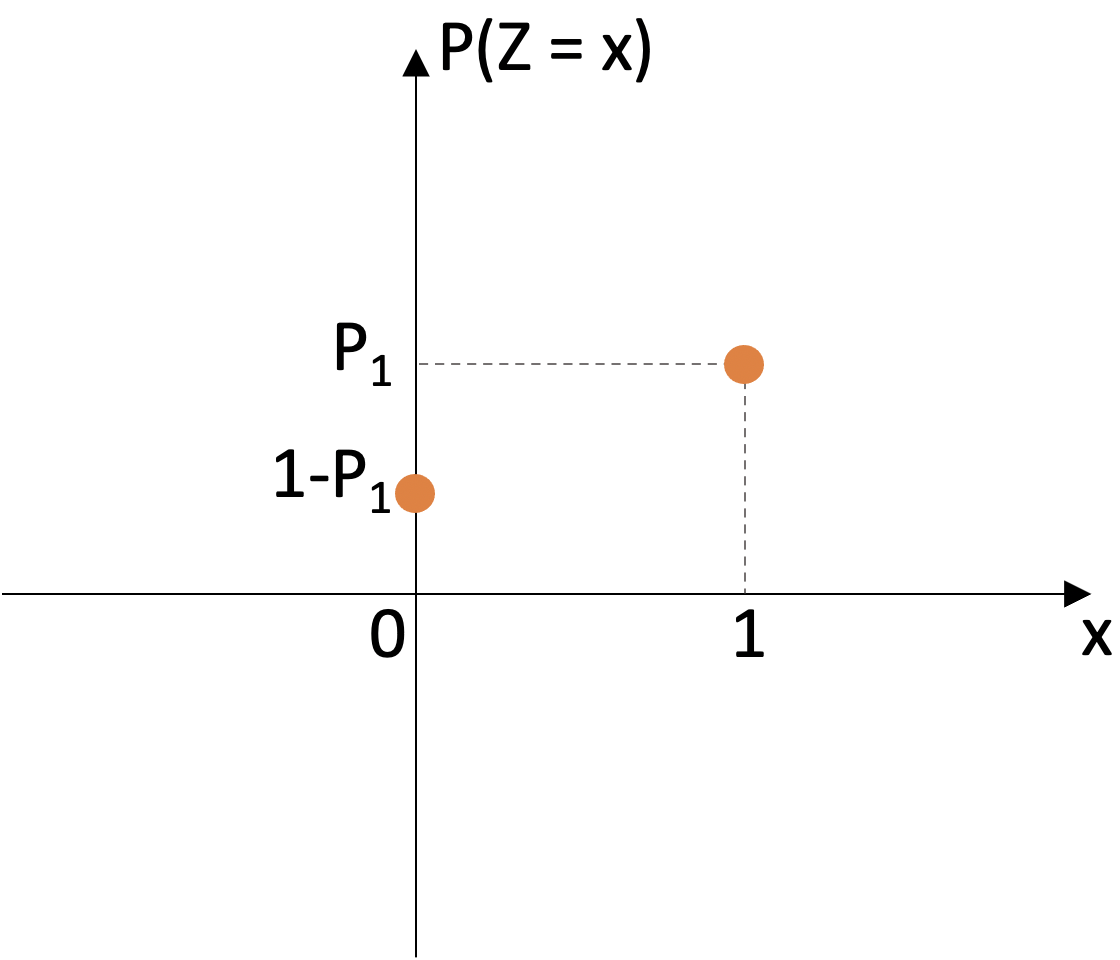}
		\caption{}
	\end{subfigure}
	\vspace{-.03in}
	\begin{subfigure}{.33\textwidth}
		\centering
		\includegraphics[width=\textwidth]{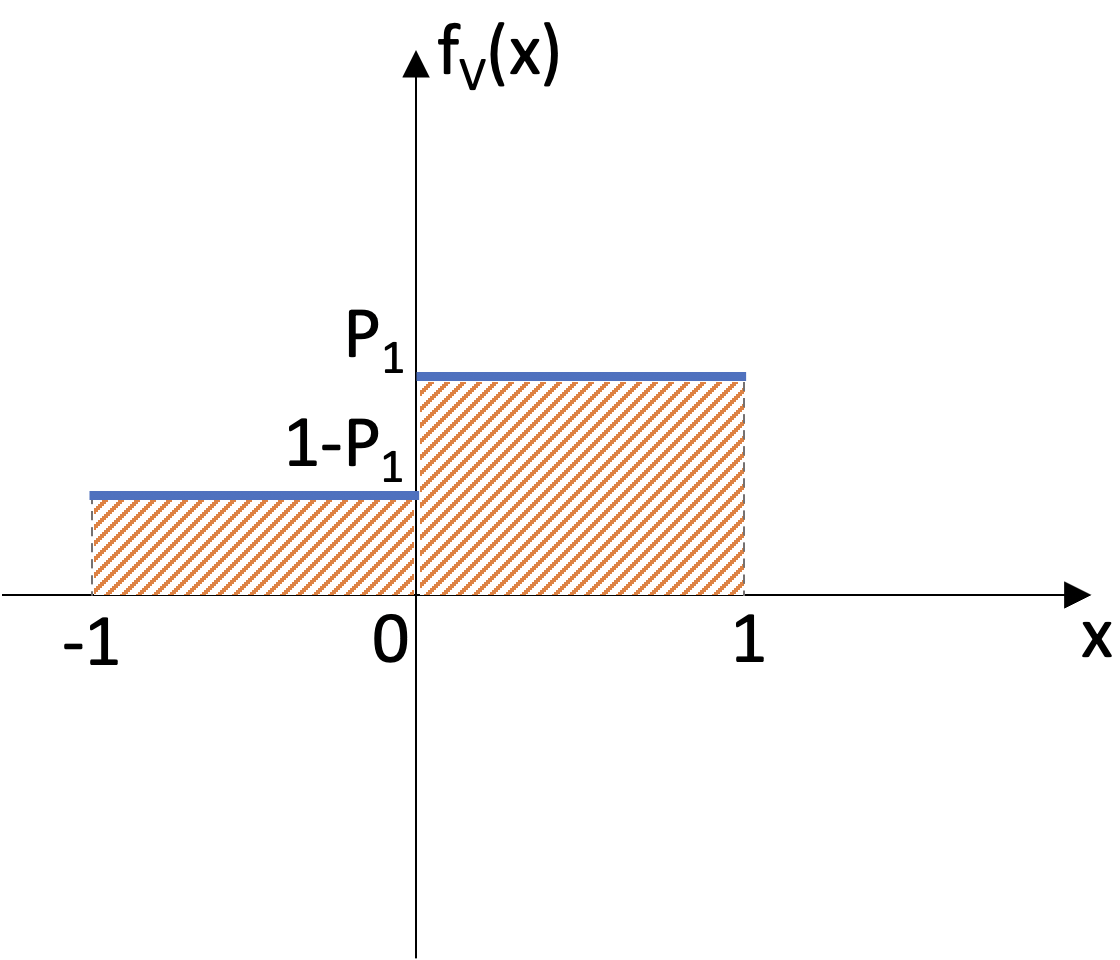}
		\caption{}
	\end{subfigure}
}

	\makebox[\linewidth][c]{
		\vspace{-.03in}
		\begin{subfigure}{.33\textwidth}\hspace{0.8in}
			\centering
			\includegraphics[width=\textwidth]{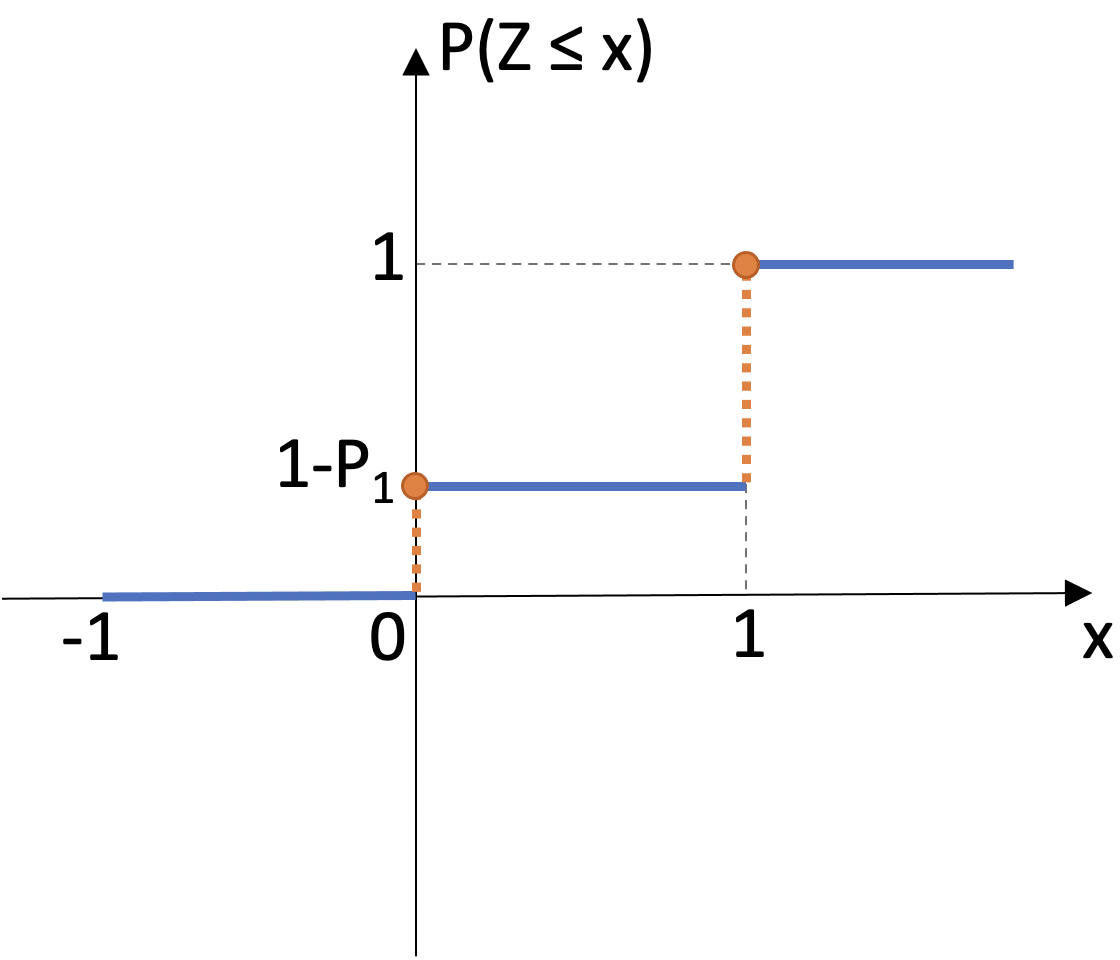}
		\caption{}
		\end{subfigure}
		\vspace{-.03in}
		\begin{subfigure}{.33\textwidth}
			\centering
			\includegraphics[width=\textwidth]{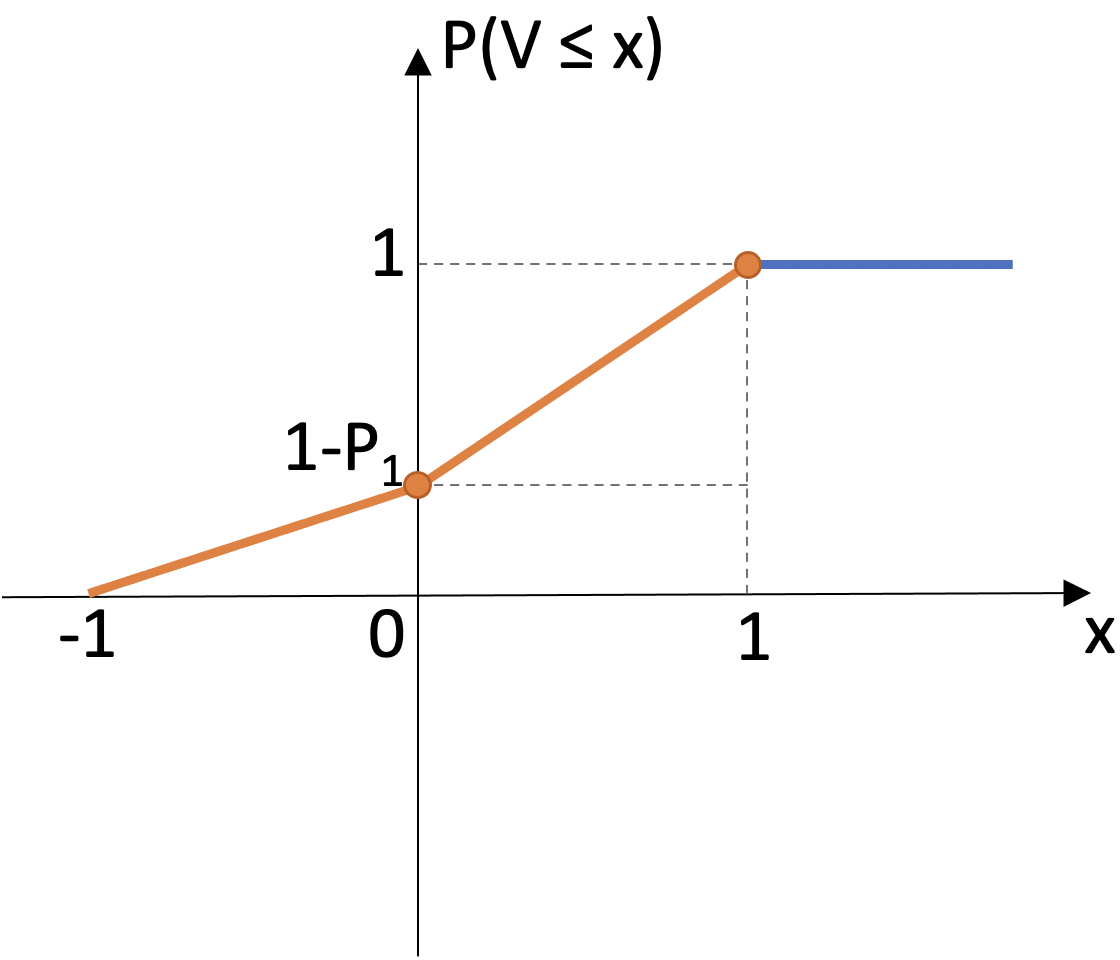}
		\caption{}
		\end{subfigure}
	}
	\caption{\textbf{An example of the univariate continualization method in DIP.} (\textbf{a}) A variable $Z$ follows a Bernoulli distribution, which takes the value 1 with probability $P_1$ and the value 0 with probability $1-P_1$. The two orange dots represent the two probability masses. (\textbf{b}) To continualize this distribution, we introduce $V=Z-U$, with $U$ being an independent random variable following a uniform distribution between 0 and 1. The two probability masses $1-P_1$ and $P_1$ are now evenly spread on intervals $(-1,0)$ and $(0,1)$ (in the orange shade), respectively. And the probability density function of this new variable $V$ is drawn in the blue lines. (\textbf{c}) The cumulative distribution function of $Z$ has two jumps -- one at 0, and the other at 1. And the heights of the jumps correspond to the probabilities of the two values. (\textbf{d}) After continualization, the cumulative distribution function of $V$ becomes continuous, where the jumps are replaced by orange lines (i.e., linear interpolants).
	}
	\label{pdf_continualization}
\end{figure}

%

\begin{table}[H]
	\begin{small}
		\begin{center}
			\begin{tabular}{ccccc}
				\hline
				\multicolumn{2}{c}{Network Reconstruction}& \multicolumn{3}{c}{$\varepsilon$}\\  \cmidrule{3-5}
				Setting	&Hold-out	&1 	&2 			&3  	\\
				\hline
				Chain Net	&15\% & 101.53 (2.82) & 101.51 (2.83)  & 101.49 (2.82) \\
				$N=200$ &25\% & 79.21 (2.21) & 79.18 (2.19) & 79.18 (2.19) \\
				$p=250$ &35\% & 75.47 (1.75) & 75.46 (1.74) & 75.45 (1.73) \\
				\hline
				Exp Decay &15\% & 42.18 (1.10) & 42.17 (1.10) & 42.16 (1.09) \\
				$N=200$ &25\% & 29.32 (1.25) & 29.31 (1.25) & 29.31 (1.25) \\
				$p=250$ &35\%  & 25.35 (1.41) & 25.35 (1.43) & 25.33 (1.43) \\
				\hline
			\end{tabular}

		\end{center}
	\end{small}
	\caption{\textbf{DIP is the only method that protects the structure of high-dimensional graphical networks.} The performance of DIP is measured by the entropy loss with its standard error in the parenthesis. Two types of networks, chain and exponential decay, are evaluated, with a sample size $N=200$ and the number of variables $p=250$. A ratio of 15\%, 25\%, and 35\% of the original sample is used as the hold-out sample to estimate the empirical distribution. The privacy factor $\varepsilon$ (i.e., the budget of privacy) ranges from 1 to 3. All of the state-of-the-art methods get infinite loss and their results are omitted.}
	\label{sim2:graphical}
	\vspace{-2mm}
	
\end{table}

\begin{table}[H]
	\begin{small}
		\begin{center}
			\begin{tabular}{cccccc}
				\hline
	\multicolumn{2}{c}{Linear Regression}		& \multicolumn{4}{c}{$\varepsilon$}\\  \cmidrule{3-6}
				Setting	&Method	&1 	&2 			&3  	& 4 \\
				\hline
				$N=200$	&NP & 0.31 (0.16) & 0.31 (0.15) & 0.31 (0.16) & 0.31 (0.15) \\ 
				$p=6$& DIP (hold 15\%) & 2.19 (1.15) & 2.16 (1.14) & 2.14 (1.13) & 2.14 (1.14) \\ 
				&DIP (hold 25\%) & 1.40 (0.77) & 1.40 (0.75) & 1.42 (0.79) & 1.40 (0.76) \\   
				&DIP (hold 35\%) & 1.13 (0.59) & 1.13 (0.61) & 1.11 (0.62) & 1.12 (0.59) \\ 
				&LRM & 62.80 (43.28) & 33.53 (21.43) & 24.53 (14.86) & 21.44 (11.91) \\ 
				&OPM & 757.16 (558.17) & 378.68 (287.99) & 241.05 (190.85) & 184.72 (136.91) \\ 		
				\hline
				$N=2000$ &NP & 0.09 (0.04) & 0.09 (0.04) & 0.09 (0.04) & 0.09 (0.04) \\ 
				$p=6$&{DIP (hold 15\%)} & 0.31 (0.15) & 0.32 (0.15) & 0.31 (0.15) & 0.32 (0.15) \\ 
				&{DIP (hold 25\%)} & 0.24 (0.12) & 0.24 (0.12) & 0.24 (0.12) & 0.24 (0.11) \\    
				&{DIP (hold 35\%)} & 0.21 (0.10) & 0.21 (0.10) & 0.21 (0.10) & 0.20 (0.10) \\ 
				&LRM & 19.55 (13.66) & 13.67 (7.94) & 12.30 (6.00) & 11.83 (4.91) \\ 
				&OPM & 650.15 (489.16) & 315.75 (240.77) & 219.46 (167.75) & 165.44 (125.28) \\ 
				\hline
				$N=200$ &NP & 0.77 (0.30) & 0.74 (0.30) & 0.76 (0.32) & 0.77 (0.33) \\ 
				$p=30$&{DIP (hold 15\%)} & 21.07 (10.64) & 20.91 (10.68) & 21.03 (10.84) & 21.05 (10.94) \\ 
				&{DIP (hold 25\%)} & 9.16 (4.41) & 9.17 (4.47) & 9.28 (4.45) & 9.21 (4.46) \\ 
				&{DIP (hold 35\%)} & 6.15 (2.84) & 6.24 (2.95) & 6.23 (2.95) & 6.21 (3.00) \\ 
				&LRM & 489.43 (370.81) & 250.98 (182.24) & 179.74 (117.89) & 138.68 (90.79) \\ 
				&OPM & 6044.5 (4611.4) & 2951.1 (2277.7) & 2021.9 (1492.6) & 1542.4 (1118.0) \\
				\hline
				$N=2000$ &NP & 0.21 (0.09) & 0.21 (0.08) & 0.21 (0.08) & 0.21 (0.08) \\ 
				$p=30$&{DIP (hold 15\%)} & 1.39 (0.59) & 1.39 (0.59) & 1.40 (0.60) & 1.39 (0.60) \\ 
				&{DIP (hold 25\%)} & 0.88 (0.38) & 0.88 (0.39) & 0.89 (0.38) & 0.88 (0.39) \\ 
				&{DIP (hold 35\%)} & 0.73 (0.32) & 0.71 (0.32) & 0.72 (0.31) & 0.72 (0.33) \\ 
				&LRM & 209.25 (147.50) & 116.93 (84.48) & 86.97 (57.60) & 74.07 (45.71) \\ 
				&OPM & 7822.7 (6055.5) & 4046.6 (3073.2) & 2595.7 (1944.0) & 2052.9 (1522.2) \\
				\hline
			\end{tabular}
			
		\end{center}
	\end{small}
	\caption{\textbf{DIP {mitigates} the trade-off between differential privacy and statistical accuracy.} Private linear regression is conducted. Averaged $L_2$-distance (standard error) between true and estimated regression coefficients are measured across different sample sizes and numbers of variables. DIP shows consistent and robust performance against the change of $\varepsilon$. NP represents the non-private benchmark.}
	\label{sim2:linear}
	
\end{table}

\begin{table}[H]
	\begin{small}
		\begin{center}
			\begin{tabular}{cc|c|c}
				\hline
				& \multicolumn{3}{c}{Dataset}\\  \cmidrule{2-4}
				& UC Salary &Bank Marketing &MovieLens\\  \cmidrule{2-4}
				Size & 252,540 & 30,488 & 25,000,095\\
				Type & Continuous & Multivariate & Discrete\\
				Dimension &1 &10 &1 \\
				Task& Mean Est. & Logistic Reg. & Collaborative Filtering\\
				\hline
				\textbf{DIP} &0.40 (0.31)  & 0.047 (0.003)  & 1.03 (4.96 $\times 10^{-4}$) \\ 
				LRM & 13.08 (9.95) & 0.311 (0.004) & 1.87 (1.03 $\times 10^{-3}$) \\ 
				OPM & 4.69 (3.44) & Infinity & 2.61 (8.23 $\times 10^{-4}$)  \\
				EXM & N/A & N/A & 1.11 (5.52 $\times 10^{-4}$) \\
				\hline
			\end{tabular}
			
		\end{center}
	\end{small}
	\caption{\textbf{DIP shows the best performance in mean estimation, logistic regression, and personalized recommendations across three benchmark datasets.} Data description, including the sample size, data type, number of variables, and the desired task, is provided on the top half of the table. The evaluation (with standard errors) of DIP and the state-of-the-art privacy protection methods (LRM, OPM, and EXM), is provided on the bottom half of the table. The evaluation metrics are the relative mean difference, the Kullback-Leibler divergence, and the root mean square error on a random $25\%$ test set for the UC salary data, the bank marketing data, and the MovieLens data, respectively. For each dataset, the same privacy protection strictness is imposed, and a smaller evaluation measure indicates a more accurate result. N/A means that a result is unavailable.}
	\label{real_logistic}
	\vspace{-3mm}
	
\end{table}

\clearpage

\section*{Appendix}

The Appendix contains the technical details of the proposed method, details about experimental setup and additional numerical studies, the connection of the proposed method with local differential privacy, generalization of the proposed method to independent but not identically distributed data, technical proofs, and additional figures and tables.

\subsection*{Technical details of the proposed method}

\subsubsection*{Convolution of Uniform and Laplace distributions}
If $U$ follows a Uniform distribution $Unif(0,1)$ and $e$ independently follows a Laplace distribution $Laplace(0,b)$, then the cumulative distribution function $G$ of $W=U+e$ is
\ba
G(x)=P(W \le x) =  \begin{cases} 
	\frac{b}{2}e^{x/b}(1 - e^{-1/b}),& x < 0 \\
	x+\frac{b}{2}e^{-x/b} - \frac{b}{2}e^{(x-1)/b}, & 0\leq x\leq 1 \\
	1-\frac{b}{2}e^{-(x-1)/b}(1 - e^{-1/b}), & x>1.
\end{cases}
\ea


\subsubsection*{Sample partitioning for the proposed method}

Consider a random sample $(Z_1,\ldots,Z_N)$ from a cumulative distribution function $F$ with either bounded or unbounded support. For DIP, we randomly partition this sample into a \emph{to-be-privatized sample}
$\bZ=(Z_1,\ldots,Z_n)$ and a \emph{hold-out sample} $\bZ^*=(Z_{n+1},\ldots,Z_{n+m})$, with $m+n=N$. 
We apply a \emph{privatization mechanism} $\mathfrak{m}(\cdot)$ to $\bZ$ to generate a privatized sample $\tilde{\bZ}=(\tilde{Z}_1,\ldots,\tilde{Z}_n)$ for release, where $\tilde{Z}_i \equiv \mathfrak{m}(Z_i)$; $i=1,\ldots,n$. 
Note that the hold-out sample is fixed once selected and is neither privatized nor released, which is used for estimating the probability distribution $F$ when it is unknown. 
{See Lemma \ref{holdout} for the privacy protection of the holdout sample.}

\subsubsection*{Univariate discrete variable}

DIP applies \eqref{ptb_c} to a discrete variable $Z_i$
after continualization. For illustration of continualization, we examine a simple example of a variable
$Z_i$ following a Bernoulli distribution $Bernoulli(P_1)$, whose distribution $F$ has a jump of size $(1-P_1)$ at $0$ and
a jump of size $P_1$ at $1$. For notational simplicity, we suppress the subscript $i$ of each variable. To continualize $F$, we introduce a new variable $V=Z-U$, with $U$ being an independent uniform 
random variable over the interval $[0,1)$. As depicted in Figure \ref{pdf_continualization}, $V$ spreads out point masses 
of $F$ at $0$ and $1$ uniformly over intervals $(-1,0]$ and $(0,1]$, respectively. 
As a result, the distribution $F_V$ of this variable $V$ becomes continuous and piecewise linear over $(-1,1)$, which also agrees with $F$ at the support of $Z$ in that $F(x)=F_V(x)$; $x=0,1$. Then, \eqref{ptb_c} applies to $V$ to yield its private version $\tilde{V}$, and subsequently, $\tilde{Z}$ is obtained by a ceiling function mapping $\tilde{V}$ to the smallest integer that is no smaller
than itself. In this way, $\tilde{Z}$ retains the original Bernoulli distribution $F$.

In general, if $Z_i$, $i=1,\ldots,n$, has non-zero probabilities at $\{a_{1}<\ldots<a_{s}\}$, then the continualization method applies with an adjustment of the domain of $U$ and the ceiling function to accommodate unequally-spaced
points $a_1,\ldots,a_s$. Specifically, if $Z_i=a_k$ for some $k \in \{1,\ldots,s\}$, we set $U_i$ as an independent uniform random variable over $[0,a_k-a_{k-1})$, with $a_{0} \equiv a_{1} -1$. Then
$V_i=Z_i-U_i$ as before becomes a uniformly distributed variable on $(a_{k-1},a_k]$. Lemma \ref{lemma1} gives an expression of 
the cdf $F_V$ of $V_i$.

\begin{customlem} {S1} \label{lemma1} 
The cumulative distribution function $F_{V}$ of $V_i$ is
Lipschitz-continuous, invertible, and $F(x)=F_V(x)$ at $x=a_k$, $k=0,\cdots,s$.
In particular, 
\be 
\label{conti}
F_{V}(x) = \frac{x-a_{k-1}}{a_{k}-a_{k-1}} P_k + \sum_{t=0}^{k-1} P_{t},  \mbox{ for } x \in (a_{k-1},a_{k}]; 
\quad P_t=P(Z_i=a_t).
\ee
\end{customlem}

On the ground of Lemma \ref{lemma1}, $H$ in \eqref{ptb_c} for a discrete variable is  
\be \label{ptbd}
\tilde{Z}_i =H(F_{V}(V_i)+e_i), \quad H= \overline{L} \circ F_{V}^{-1}\circ G,
\ee
where 
$\overline{L}(x) \equiv \inf\{a_{k}: a_{k}\ge x\}$ is a generalized ceiling function, and hence $\tilde{Z}_i$ follows 
the original distribution $F$.

Note that, for a categorical variable with $s$ categories, \eqref{ptbd} is not applicable, but we can treat it with \eqref{multi} as an $(s-1)$-dimensional multivariate variable.

\subsubsection*{Univariate mixed variable}
DIP \eqref{ptbd} extends to most mixed distributions consisting of both continuous and discrete components,
such as mixture distributions with a finite number of components. 

For a mixed cumulative distribution function $F$, we need to treat jump discontinuities. By non-decreasingness of $F$
and Darboux-Froda's Theorem (Theorem 4.30 of \cite{rudin1964principles}), the set of jump discontinuities 
is at most countable, denoted by jump discontinuous points $\{a_0, a_{\pm 1},\ldots,a_{\pm s}\ldots\}$, 
where $a_k < a_{k+1}$ for any $k$, and $a_0$ is defined as the first non-negative jump such that $a_{-1} < 0 \le a_{0}$.

Now we modify the continualization method in \eqref{ptbd}. 
Note that, unlike in the discrete case, it is possible that $P(a_{k-1} < Z_i < a_{k})>0$ for certain $k$'s since $Z_i$ contains both continuous and discrete components. To circumvent this difficulty, we evenly spread the probability mass at each $a_{k}$ by ``squeezing in'' a unit interval to its left,
instead of evenly spreading across $(a_{k-1},a_{k}]$ as in \eqref{ptbd}. See Figure \ref{mixed_figure} for an illustration.
Then $Z_i$ is continualized by $V_i=Z_i+k-U_i$ when $Z_i = a_{k}$ and $V_i=Z_i+k$ when $Z_i \in (a_{k},a_{k+1})$;
$k=0,\pm 1,  \pm 2, \ldots,$
where $U_i$ follows a Uniform distribution on $(0,a_k-a_{k-1}]$ given $Z_i=a_k$ and the cdf of $V_i$ is continuous. Now, we apply \eqref{ptb_c} to $V_i$ to 
yield $\tilde{V}_i$ and then apply a generalized ceiling function $\olL(x)$ to retain the distribution of $Z_i$, that is,
\be \label{cont_mixed}
L_0(\tilde{V}_i) \equiv \begin{cases} 
	\olL(\tilde{V}_i-k), & \mbox{if $\tilde{V}_i \in [a_{k}+k-1,a_{k}+k]$, }k=0,\pm 1,\cdots\\
	\tilde{V}_i-k,& \mbox{if $\tilde{V}_i \in (a_{k}+k,a_{k+1}+k)$, } k=0,\pm 1,\cdots 
\end{cases}
\ee
where $\olL(x)=a_{k}$ for $x \in [a_{k}-1,a_{k}]$. In summary, the DIP mechanism for a mixed variable $Z_i$ becomes 
\be \label{ptb_mixed}
\tilde{Z}_i= H\big(F_{V}(V_i)+e_i\big), \quad H=L_0 \circ F_{V}^{-1}\circ G,
\ee
where $L_0$ is defined in \eqref{cont_mixed},
and $\tilde{Z}_i$ continues to follow the distribution $F$ of $Z_i$ and is
$\varepsilon$-differentially private. A formal statement of the results is included in Theorem \ref{dp_multivar}.



\subsubsection*{Univariate variable with an unknown distribution}

When $F$ is unknown, we replace it with a good estimate $\hat{F}$, such as an empirical distribution 
function (edf) based on an independent hold-out sample of $F$. 
{For an edf $\hat{F}$, we continualize it as in \eqref{conti} and then apply \eqref{ptbd} to achieve privatization.}

We continualize edf $\hat{F}$ based on a hold-out sample $\bZ^*$ of size $m$ to yield
a continuous cdf $\hat{C}$.
Specifically, we create a new sample $\bV^*$ which is equal to $\bZ^*$ 
if the original $F$ is continuous, and is equal to a continuous version of $\bZ^*$ if $F$ is discrete or mixed, as 
defined in the univariate discrete and mixed cases.
Then a continuous version of $\hat{F}$, namely $\hat{C}$, is created:
\be \label{continualized_edf}
\hat{C}(x) = \frac{1}{m}\cdot\frac{x-d_{k-1}}{d_k-d_{k-1}} + \frac{k-1}{m}, \mbox{ for } x \in (d_{k-1},d_k]; \quad k=1,\ldots,m
\ee
where 
$d_k=V_{(k)}^*$ is the $k$th order statistic of $\bV^*$. 
On this ground, the privatized $Z_i$ becomes  
\be \label{ptb_empirical2}
\tilde{Z}_i = H\big(\hat{C}(V_i)+e_i\big), \quad H=L \circ \hat{C}^{-1}\circ G, 
\ee
where $V_i$ is defined in Lemma \ref{lemma1}, and 
$L$ is an identity function as in \eqref{ptb_c} if $F$ is continuous, $L=\bar{L}$ as in 
\eqref{ptbd} if $F$ is discrete, and $L=L_0$ as in   
\eqref{cont_mixed} if $F$ is mixed.

\subsubsection*{Multivariate variable with unknown distribution}


For an unknown multivariate cdf $F$, we generalize \eqref{multi} to an empirical distribution $\hat{F}$, 
where $\hat{F}(\bx)=m^{-1}\sum_{j=1}^m\prod_{l=1}^{p}\mathds{1}_{[Z^*_{jl},\infty)}(x_l)$
given a hold-out sample $\bZ^*$, and $\mathds{1}_A(t)$ is an indicator with $\mathds{1}_A(t)=1$ if 
$t\in A$ and $0$ otherwise. 
{Similar to the univariate case, we have to create a continuous version $\hat{C}$, such that \eqref{multi} can be applicable to each conditional distribution.}

Towards this end, we propose a multivariate continualization method. 
We first create a continuous version $\bV^*$. For each variable of $\bV^*$, it is equal to the corresponding variable of $\bZ^*$ if $\bZ^*$ is continuous, or its continuous version otherwise. Then 
the basic idea of multivariate continualization is to split the support of $\hat{F}$ into a grid of small $p$-dimensional cubes and evenly spread the probability mass of each observation across the cube on the immediate bottom-left of the observation. 
For any $\bx=(x_1,\ldots,x_p)$ in the support of $\hat{F}$, we identify the cube it belongs to
so that the corresponding $\hat{C}(x_1)$ and $\hat{C}(x_l|x_1,\ldots,x_{l-1})$; $l=2,\ldots,p$, can be computed accordingly.

An example in Figure \ref{new_2d_continualization} illustrates the
continualization process of an empirical conditional distribution, where
an hold-out sample $\bZ^*$ has five observations $(d_{11},d_{23})$, $(d_{12},d_{21})$, $(d_{13},d_{25})$, $(d_{14},d_{22})$, and $(d_{15},d_{24})$. Of particular
interest is the empirical distribution $\hat{F}(x_1,x_2)$ built on $\bZ^*$, as well as
their values, for example, $\hat{F}(d_{11},d_{23})=0.2$ at a jump point $(d_{11},d_{23})$ and 
$\hat{F}(d_{11}/2,d_{23})=0$ at a non-jump point $(d_{11}/2,d_{23})$,
as depict in Figure \ref{new_2d_continualization}(a). As displayed in
Figure \ref{new_2d_continualization}(b), we evenly spread
each of the 5 probability masses in $\bZ^*$  across the corresponding bottom-left rectangles, which forms a new distribution. 
On this ground, we construct a smooth version $\hat{C}$ for this new distribution, which matches
up the values of $\hat{F}$ at jump points, for example, $\hat{C}(d_{11},d_{23})=
\hat{F}(d_{11},d_{23})=0.2$. However, $\hat{C}(d_{11}/2,d_{23})=0.1 \neq \hat{F}(d_{11}/2,d_{23})$ for a non-jump point of $\hat{F}$ due to the spread mass in 
the leftmost cube. For the $i$th observation $\bZ_i=(Z_{i1},Z_{i2})$ of the to-be-privatized sample, we privatize $Z_{i1}$ into $\tilde{Z}_{i1}$ following \eqref{ptb_empirical2}. Suppose $\tilde{Z}_{i1}=\tilde{z}_{i1} \in (d_{12},d_{13}]$, then Figure \ref{new_2d_continualization}(b) shows the domain of the smoothed conditional distribution of $Z_{i2}$ given $\tilde{Z}_{i1}=\tilde{z}_{i1}$, that is, $\hat{C}(x_2|x_1=\tilde{z}_{i1})$, as represented by the dashed line. The actual curve of the smoothed cdf $\hat{C}(x_2|x_1=\tilde{z}_{i1})$ is shown in Figure \ref{new_2d_continualization}(c), which has non-zero probability only in $(d_{24},d_{25}]$, corresponding to the spread probability mass in Figure \ref{new_2d_continualization}(b). An
analytic expression of the conditional distribution is 
\ba
\hat{C}(x_2|x_1=\tilde{z}_{i1})=
\begin{cases}
	1 & \text{if } x_2 > d_{25}\\
	\frac{x_2-d_{24}}{d_{25}-d_{24}} & \text{if } x_2 \in (d_{24},d_{25}]\\
	0 & \text{if } x_2 \le d_{24},
\end{cases}
\ea
and $\hat{C}(x_2|x_1)=0$ if $x_1 \notin (0,d_{25}]$. 

To introduce a general form of the conditional cdf 
$\hat{C}(x_l|x_1,\ldots,x_{l-1})$ for 
any $\mathbf{x}$, let $d_{jl}$ be the $j$th order statistic of the $l$th variable of $\bV^*$; $j=1,\ldots,m$, $l=1,\ldots,p$.
Note that if there exists $q \in \{1,\ldots,m\}$ such that $(x_1,\ldots,x_{l-1})$ is within the smallest cube on the 
immediate bottom-left of $(V^*_{q1},\ldots,V^*_{q,l-1})$, then such a $q$ in the hold-out sample is unique 
since the order statistics are non-identical almost surely. If the corresponding $V^*_{ql}$ be the $j$th order statistic of the $l$th variable of $\bV^*$, that is, $V^*_{ql}=d_{jl}$, then,
\be \label{multi_cond}
\hat{C}(x_l|x_1,\ldots,x_{l-1}) =
\begin{cases}
	1 & \text{if } v_l > d_{jl}\\
	\frac{v_l-d_{j-1,l}}{d_{jl}-d_{j-1,l}} & \text{if } v_l \in (d_{j-1,l}, d_{jl}]\\
	0 & \text{if } v_l \le d_{jl},
\end{cases}, \mbox{ }l=2,\ldots,p,  
\ee
and $\hat{C}(x_l|x_1,\ldots,x_{l-1}) = 0$ if such $q$ does not exist.

On the ground of \eqref{multi_cond}, we apply $\hat{C}(x_l|x_1,\ldots,x_{l-1})$ to the sequential mechanism \eqref{multi}. 
To privatize $\bZ=(\bZ_1,\ldots,\bZ_n)$ given $\hat{C}$, we first convert $\bZ$ into $\bV=(\bV_1,\ldots,\bV_n)$, same as the creation of $\bV^*$, and followed by an application of \eqref{multi} to each $\bV_i$ with each $\fm_{l}(\cdot)$: 
\be \label{cond_priv}
\tilde{V}_{il} \equiv \fm_l(V_{il})=H\big(\hat{C}(V_{il}|V_{i1},\ldots,V_{i,l-1})+e_i\big), \quad H=\hat{C}^{-1}(\cdot|\tilde{V}_{i1},\ldots,\tilde{V}_{i,l-1})\circ G(\cdot)
\ee
to yield its privatized version $\tilde{\bV}_i$. Then, we apply $L_l(\tilde{V}_{il})$ to convert $\tilde{\bV}_i$ back to $\tilde{\bZ}_i$
for each $l$ , where $L_l(\cdot)$ is the ceiling function corresponding to the $l$th variable of $\tilde{\bV}_i$; $l=1,\ldots,p$.

%
%
%
%
%

\subsection*{Experimental setup and additional numerical studies}

\subsubsection*{Univariate discrete distributions}

For the univariate case, we assume the true distribution is known, and thus a hold-out sample is not necessary.

We first consider private sample mean estimation. We use a discrete variable with a known
distribution of bounded or unbounded support.
A random sample is generated, with the size being 1000, and the distribution (with the parameter value in the parenthesis) is a Bernoulli distribution $Bernoulli(0.1)$, a Binomial distribution $Binomial(5,0.5)$, 
a Poisson distribution $Poisson(3)$, or a Geometric distribution 
$Geometric(0.2)$. Here we privatize the simulated data and
measure the estimation accuracy by the difference between the true mean and 
the private mean. The privacy factor ranges from 1 to 4, indicating strict privacy protection.

Three privatization methods are examined via simulations. For DIP, we apply \eqref{ptbd}. For LRM, we add Laplace noise following $Laplace(0,1/\varepsilon)$. For OPM, we apply the private mean 
in Section 3.2.1 of \cite{duchi2018minimax} under the assumption of the bounded $k$-th moment with $k=\infty$, which provides the most accurate result. 
For EXM, the discrete sampling distribution for privatization depends on a quality score function and a sensitivity measure (i.e., the range of the distribution),  in which we use the frequency of each discrete value as the quality score and approximate the sensitivity by the maximum value of a sample. 
In what is to follow, DIP is compared with its non-private counterpart (NP) when available, in which 
no privatization applies. The study is repeated 1000 times.

As indicated in Table \ref{sim2_para_est}, DIP again delivers the lowest estimation
error as well as the standard deviation over its competitors LRM, OPM, and EXM, across all settings. The amount of improvement of DIP over LRM, OPM, and EXM ranges from 34.8\% to 1874.7\%.  
Moreover, DIP is insensitive to a change of $\varepsilon$. Interestingly,
EXM performs better than LRM and OPM in estimation, while all the three methods perform
worse as the value of $\varepsilon$ decreases, where EXM is least impacted 
by $\varepsilon$.

In Figure \ref{fig:ParaEst}, we evaluate each method with a fixed $\varepsilon=1$ and a varying sample size from 100 to 2000. DIP performs nearly identically
to its non-private counterpart which uses the original data due to its distribution preservation property, and enjoys the fastest convergence of the error rate as the sample size increases. Meanwhile, DIP delivers the lowest estimation error among its competitors LRM, OPM, and EXM across all settings. This simulation demonstrates the importance of distribution preservation as evident by the deterioration amount of LRM, OPM, and EXM over DIP, ranging from 41.2\% to 1874.7\% as a result of alternation of the original distribution.

\subsubsection*{Univariate continuous distributions}

This simulation examines the distributional distortion of privatization
by DIP, LRM, and OPM, as measured by the Kolmogorov-Smirnov distance 
between the true and the empirical distributions, as a function
of the privacy factor and the sample size. 

A random sample $Z_1,\ldots,Z_n$ is
generated from a Uniform distribution $Unif(0,1)$, a Beta distribution $Beta(2,5)$, a Normal distribution $N(0,1)$, 
and an Exponential distribution $Exp(1)$, with $n=1000$.
For each $\varepsilon$, we perform privatization by DIP, LRM, and OPM to
yield a privatized sample $\tilde{\bZ}=(\tilde{Z}_1,\ldots,\tilde{Z}_n)$,
while $\tilde{\bZ}=\bZ$ is un-privatized for NP. For DIP, we apply 
\eqref{ptb_c} with the known $F$. Note that LRM and OPM require a bounded 
domain of $F$, which is not met by $N(0,1)$ and $Exp(1)$. For LRM, we use $[-\max_{i=1,\ldots,n}(|Z_i|),\max_{i=1,\ldots,n}(|Z_i|)]$ and $[0,\max_{i=1,\ldots,n}(Z_i)]$ to approximate the 
unbounded support of $N(0,1)$ and $Exp(1)$, respectively. 
For OPM, we use $\sqrt{n}\max_{i=1,\ldots,n}(|Z_i|)$ to approximate
the true radius of a $l_2$-ball, following the privacy mechanism for a 
$l_2$-ball in Section 4.2.3 of \cite{duchi2018minimax}. 

As indicated in Table \ref{sim1}, DIP performs the best across 16 situations by comparing with OPM and LPM and performs nearly the same as the oracle method NP. The amount of improvement of DIP over LRM ranges from $(129.55-26.91)/26.91=381\%$ to $(439.61-27.62)/27.62=1492\%$,
while that over OPM is from $(507.38-27.42)/27.42=1750\%$ to $(510.14-26.99)/26.99=1790\%$. Moreover, DIP is insensitive to a change of $\varepsilon$,  whereas LRM's and OPM's performance deteriorates when a smaller value of $\varepsilon$ imposes a stricter differential privacy policy. This aspect is due to the distribution-preservation property of DIP's privatized data $\tilde{\bZ}$ regardless of the value of $\varepsilon$ when privatizing for a known $F$. As suggested by Theorem \ref{dp_multivar} and the related discussions, DIP entails no loss of statistical accuracy with a known distribution $F$.


\subsubsection*{Gaussian graphical models}

In subsequent experiments, the true data distribution is assumed to be unknown and thus a hold-out sample is used.

Consider privatized Gaussian graphical model with an unknown multivariate normal distribution 
$F$. Simulations are performed in low-dimensional situations with $N=2000$, where a sample
of size $N=n+m$ includes a hold-out sample and a sample to be privatized, and $p=5, 15$
and the privacy factor $\varepsilon=1,2,3,4$. Two types of graph networks are examined, including the chain and exponential
decay networks. Let $\Omega$ be the true precision matrix, that is,
the inverse of the covariance matrix of the normal distribution. For the chain network,  $\Omega$ is a tri-diagonal matrix corresponding to the first-order autoregressive structure. 
Specifically, the $(k,l)$th element of $\Omega^{-1}$ is $\sigma_{kl}=\exp(-0.5|\tau_k-\tau_l|)$, where $\tau_1<\cdots < \tau_p$ and $\tau_l - \tau_{l-1}$ follows a Uniform distribution $Unif(0.5,1)$; $l=2,\ldots,p$. 
For exponential decay network, the $(k,l)$th element of 
$\Omega$ is $\omega_{kl}=\exp(-2|k-l|)$. Clearly, $\Omega$ is sparse 
in the first case but not in the second. 

Based on privatized data,  we estimate $\Omega$ using 
graphical Lasso \citep{friedman2008sparse} in 
the “glasso” R-package with 5-fold cross-validation. The estimated precision matrix 
$\hat{\Omega}$ is evaluated by the entropy loss \citep{lin1985monte} by comparing 
with the true precision matrix $\Omega$, denoted by $EL(\Omega,\hat{\Omega})=\mbox{tr}(\Omega^{-1}\hat{\Omega}) - \log |\Omega^{-1}\hat{\Omega}| -p$.
The rest of the setup is identical to the graphical model described in the main text.

For LRM and OPM,
although it remains unclear how to generalize them to the multivariate case, we suggest a naive generalization, 
respectively referred to as naive LRM (NLRM) and naive OPM (NOPM), 
to privatize each variable individually while ignoring 
the multivariate dependence structure.
The approximation used for unbounded support is the same as the one described in the previous example.
For NLRM and NOPM, the entire sample is released, and each variable is privatized independently under a privacy factor $\varepsilon/p$.

As suggested by Table \ref{sim2:graphical}, DIP has the lowest error of estimating the precision matrix in low-dimensional situations. Its improvement over competitors ranges from 18437\% to 91518\% in terms of entropy loss. Moreover, it yields a much smaller standard error than NLRM and NOPM do. Moreover, DIP performs similarly to NP, indicating that its distribution preservation plays a crucial role in retaining
the dependency structure. The closeness between DIP and NP
is because of a small error in estimating the unknown multivariate cdf $F$. Moreover, DIP is insensitive to the value of $\varepsilon$ than NLRM and NOPM. 
In summary, DIP's distribution preservation property becomes more critical to recovering multivariate structures, which interprets the phenomenon of sizable differences between DIP and its competitors in simulations.

\subsubsection*{Linear regression}

We provide more technical details of the linear model simulation in the main text. The response variable vector
$\bY=(Y_1,\cdots,Y_N)$ is generated according to $\bY=\bX\bbeta+\mathbf{e}$, where $e_i$ follows $N(0,1)$, $\bX$ is a $N \times p$ design matrix,
and $\bbeta$ is a $p$-dimensional vector of coefficients. In simulations,
$N=200$ or $2000$, the privacy factor $\varepsilon=1,2,3,4$,
$\bbeta_{p \times 1}=(1,1,\ldots,1)$ with $p=6$ or $30$, and each 1/3 of
$\bX$'s columns follow independent a normal distribution $N(0,10^2)$, a Poisson distribution $Poisson(5)$, and
a Bernoulli distribution $Bernoulli(0.5)$, respectively.
In this example, we estimate the regression coefficient
vector $\bbeta$ based on the privatized release data
and measure the estimation accuracy by the $L_2$-norm between the estimated and true parameter vectors.

For DIP, we split the data into hold-out and to-be-privatized samples with a splitting
ratio of 15\%, 25\%, and 35\%. Then we apply Algorithm \ref{alg:multivar} to privatize
variables of the to-be-privatized sample in a random order to examine DIP's invariant property
of the sequential order in Theorem \ref{consistency_multivar}.
For LRM and OPM, we adopt the univariate case as in the graphical model case, and utilize the independence of columns of $\bX$
to convert multivariate privatization to the univariate case. In particular,
we privatize $\bX$ to yield $\tilde{\bX}$, followed by privatizing $\bY$ to yield
$\tilde{\bY}$ given $\tilde{\bX}$ by privatizing the residuals after
fitting $\bY$ on $\tilde{\bX}$. Then we regress $\tilde{\bY}$ on $\tilde{\bX}$ to
produce an estimated regression parameter vector.
Note that LRM and OPM utilize the additional information -- independence among columns of $\bX$, whereas
DIP does not.
For all methods, each variable is privatized under a privacy factor $\varepsilon/(p+1)$. The study is repeated 1000 times.

\subsubsection*{The University of California salary data} \label{sec:salary}

The University of California (UC) system collects annual salaries of $N=252,540$ employees, 
including faculty, researchers, and staff. Data are available at \url{https://web.stanford.edu/~jduchi/publications.html}.
For the Year 2010, the average salary of all employees is \$39,531.49 with a 
standard deviation \$53,253.93. The data is highly right-skewed, with
the 90\% quantile \$95,968.12 and the maximum exceeding two million dollars.

For this dataset, we apply each mechanism to estimate the $\varepsilon$-differentially private mean UC salary.
One important aspect is to contrast the privatized mean with the original mean
\$39,531.49 to understand the impact of privatization on statistical accuracy
of estimation. Three privatization mechanisms are compared, including
DIP, LRM, and OPM. For DIP, we hold out $15\%$, $25\%$, or $35\%$ of the sample, and apply Algorithm \ref{alg:multivar}. For OPM, we follow the private mean estimation function described in Section 3.2.1 of \cite{duchi2018minimax} and choose the number of moments $k=20$ and the moment as the one closest to $3$ to optimize its performance. The above process,
including privatization, is repeated 1000 times.

As indicated in Table \ref{real1_mean}, DIP delivers the most accurate mean salary estimation under differential privacy. The amount of improvement on LPM and OPM is in the range of 405\% to 3533\%.
By comparison, LRM and OPM yield a large estimation error.
Note that the performance is attributed primarily to the distribution-invariant property that LPM and OPM do not possess. Moreover, the cost of stricter privacy protection is little for DIP. When $\varepsilon$ decreases from 4 to 1, DIP's relative error increases only by 35\%, 33\%, and 24\%, given $15\%$, $25\%$, and $35\%$ of the sample held out, respectively. 
By comparison, those of LRM and OPM increase by 288\% and 151\%, respectively. This
is a result of the impact of the high privatization cost of LPM and OPM on statistical accuracy.
In summary, the distribution preservation property is critical to maintaining
statistical accuracy in downstream analysis.

	%
		%
%

\subsubsection*{The Portuguese bank marketing campaign data}

This bank marketing campaign intends to sell long-term deposits to potential clients through phone conversations. During a phone call, an agent collects a client's data, past contact histories, and if the client is interested in subscribing to a term deposit (yes/no). The marketing campaign data are available at \url{http://archive.ics.uci.edu/ml/datasets/Bank+Marketing#}.
Here our goal is to conduct privatized logistic regression and examine the statistical accuracy change after privatization.

The data includes a response variable and 9 explanatory variables. The response variable is whether a client has an interest in subscribing a long-term deposit. Nine explanatory variables include clients' age (numeric), employment status (yes/no, where ``no'' includes ``unemployed'', ``retired'', ``student'', or ``housemaid''), marital status (``single'', ``married'', and ``divorced'', which is coded as two dummy variables with ``married'' being the reference level), education level (``illiterate'', ``4 years'', ``6 years'', ``9 years'', ``professional course'', ``high school'', and ``university'', labeled as 0 to 6, respectively), default status (yes/no), housing loan (yes/no), personal loan (yes/no), client's device type (``mobile'' or ``landline''), and the total number of contacts regarding this campaign.

For NLRM, Laplace noise is added to each variable independently. For NOPM, we conduct private logistic regression following the private estimation of private generalized linear models in Section 5.2.1 of \cite{duchi2018minimax}. The privatization process is repeated 1000 times.

As shown in Table \ref{real2_logistic}, parameter estimation by DIP yields a small value of the Kullback-Leibler divergence -- less than $5 \times 10^{-2}$. Moreover, its performance is insensitive to the private factor $\varepsilon$, permitting the low cost of strict privacy protection, which is ensured by the distribution-invariant property, c.f., Theorem \ref{consistency_multivar}. DIP performs slightly better if we use more data in the hold-out sample for empirical cdf estimation. In contrast, the performance of NLRM is at least five times worse than that of DIP, and the results provided by NOPM are infinity since an estimated probability of 1 or 0 exists across all settings.

		%
%

\subsubsection*{The MovieLens data}

Privatization is critical to protect personal information for released
consumer data. In many situations, simply removing users' identities is not
adequate as suggested in \citep{demerjian2007rise,narayanan2008robust},
in which political preferences and other sensitive information 
can be leaked even if the IDs are masked. To overcome this difficulty,
one needs to privatize movie ratings. 

For illustration, consider the most recent MovieLens 25M dataset collected by GroupLens Research \cite{harper2015movielens} available at \url{https://grouplens.org/datasets/movielens/25m}. This dataset contains $N=25,000,095$ movie 
ratings in $\{0.5,1,1.5,\ldots,5\}$, collected from 162,541 users over 59,047 
movies between 1995 and 2019. 

To investigate the effect of data privatization on the prediction accuracy 
for movie ratings, we randomly split the movie ratings into a 75\% training set and a 25\% test set. Moreover, we treat all ratings, for instance, ratings from all users assigned to all movies, from a single distribution. For DIP, we randomly select
15\%, 25\%, or 35\% of the training sample as a hold-out sample and apply Algorithm \ref{alg:multivar}. Then we apply the univariate privatization method with unknown true distribution in (\ref{ptb_empirical2}) to privatize the to-be-privatized sample. Note that only the privatized sample is used in the recommender algorithm. 

For other methods, the entire training set is privatized and trained after privatization.
For LPM, Laplace noises are added to the raw ratings, and the noised ratings are rounded to the nearest number in $\{0.5,1,1.5,\ldots,5\}$. For OPM, the method described in the example of univariate continuous distributions applies with the radius approximated as $4.5\sqrt{N}$. For EXM, the discrete sampling distribution for privatization depends on a quality score function and a sensitivity measure (i.e., the range of the distribution),  in which we use the frequency of each discrete value as the quality score and approximate the sensitivity by the maximum value of a sample. The privacy factor $\varepsilon$ for all methods is set to be 1.
Then we train a matrix factorization model \citep{funk2006netflix} based 
on privatized ratings, which is a prototype collaborative filtering method 
for movie recommendation. We use the ``recosystem'' package in R with cross-validation for tuning. The evaluation metric is the root mean square error
on the original test set, which is averaged over 50 random partitions of training,
test, and hold-out samples.

As indicated in Table \ref{real_logistic} and Table \ref{SVD}, DIP delivers
the desirable performance across all settings and even
performs better with a smaller hold-out sample. This is in contrast to the 
previous studies in which a larger hold-out sample usually yields better
parameter estimation. 
Moreover, LRM and EXM have improved performance when $\varepsilon$ is large, and
LRM performs similarly to DIP's when $\varepsilon=4$.

		%
%

\subsection*{Local differential privacy}

Definition \ref{DP} protects data owners' privacy after aggregating all data points to a data server and then applying privatization. In many scenarios, however, data owners may not even trust the data server. To further protect data owners' privacy from the server, one may consider data privatization at each owner's end before aggregation. And the notion of $\varepsilon$-local differential privacy is introduced \citep{evfimievski2003limiting,kasiviswanathan2011can}. 
\begin{customdef}{S1} 
	\label{LDP}
	A privatization mechanism $\fm(\cdot)$ satisfies $\varepsilon$-local 
	differential privacy if
	\ba
	\sup_{z_i,z_i'}\sup_{B_i}\frac{P\big(\fm(Z_i) \in B_i|Z_i=z_i\big)}{P\big(\fm(Z_i) \in B_i|Z_i=z_i'\big)} \le e^{\varepsilon},
	\ea
	where $B_i$ is a measurable set and $\varepsilon \ge 0$ for
	$i=1,\ldots,n$. The ratio is defined as 1 if the numerator and denominator are zero.
\end{customdef}

\begin{customlem}{S2} \label{trans_LDP}
	If $\tilde{Z}_i = \fm(Z_i)$ satisfies $\varepsilon$-local differential privacy, then $\zeta(\tilde{Z}_i)$ is also $\varepsilon$-locally differentially private for any measurable function $\zeta(\cdot)$.
\end{customlem}

Lemma \ref{trans_LDP} establishes a parallel result of Lemma 2.6 of \cite{wasserman2010statistical}.
Based on Lemma \ref{trans_LDP}, we may generalize the DIP mechanism  
\eqref{multi} to local differential privacy. 
Further investigation is necessary to understand DIP under local differential privacy.

\subsection*{DIP for Independent but not Identically Distributed Data}

DIP can be generalized to  an independent but not identically distributed 
sample to be differentially private while approximately preserving the 
distribution and thus any dependence structure therein. Specifically,
given an original sample $(Z_1,\ldots,Z_n)$, with $Z_i \sim F_i$; $i=1,\ldots,n$ independently. Furthermore, when $F_i$ is unknown, assume that $J$ repeated 
measurements are available for each individual $i$, or $Z_{ij}$; $j=1,\ldots,J$. 
Then, we use a random subset of each 
individual's data as a hold-out sample. For example, we use $Z_{i1},\ldots,Z_{ij_0}$, $1 \le j_0 \le J$, as the hold-out sample to construct the empirical CDF 
$\hat{F}_i$; $i=1,\cdots,n$, and privatize and release $Z_{i,j_0+1},\ldots,Z_{iJ}$ by  
\ba
\tilde{Z}_i = \hat{F}_i^{-1} \circ G (\hat{F}_i(Z_i) + e_i),
\ea
where $\tilde{Z}_i$ satisfies differential privacy.
As in a fixed effect model, DIP may not be applicable (when $J=1$) or accurate for preserving
the distribution of $F_i$'s when $J$ is small, although it is differentially private.

\subsection*{Proofs}


\subsubsection*{Proof of Lemma \ref{rep_queries}}

Without loss of generality, assume that $i_0=1$. 
For any $\mu_1 \neq \mu_0$, the Neyman-Pearson Lemma states that the most powerful test is to reject the null hypothesis at: 
$B=\{(\tilde{\bZ}^{(1)},\ldots,\tilde{\bZ}^{(M)}): P(\tilde{\bZ}^{(1)},\ldots,\tilde{\bZ}^{(M)}|\bz) > \nu P(\tilde{\bZ}^{(1)},\ldots,\tilde{\bZ}^{(M)}|\bz')\}$,
where $\nu$ is the critical value such that $P(B|Z_1=\mu_0)=\gamma$, the level of significance, and $\bz= (\mu_1,z_2,\ldots,z_n)$
and $\bz'= (\mu_0,z_2,\ldots,z_n)$ are two adjacent realizations of $\bZ$. By Definition \ref{DP}, $P(\tilde{\bZ}^{(j)}|\bz) \le e^{\varepsilon} P(\tilde{\bZ}^{(j)}|\bz')$; $j=1,\cdots,M$. By conditional independence of $\tilde{\bZ}^{(1)},\ldots,\tilde{\bZ}^{(M)}$ given $\bZ$, $P(B|\bz) 
\le 
e^{M\varepsilon} P(B|\bz')$. Then the power 
of this test is  
\ba
P(B|Z_1=\mu_1) &=& \int P(B|\bz) dP(z_2,\ldots,z_n)
\le e^{M\varepsilon}  \int P(B|\bz') dP(z_2,\ldots,z_n)\\
&=& e^{M\varepsilon} P(B|Z_1=\mu_0) = \gamma e^{M\varepsilon}.
\ea
This completes the proof. 


\subsubsection*{Proof of Lemma \ref{direct_ptb}}

For an arbitrary $\bt_0$, the probability density (or mass) of $\fm(\bZ_i) = \bt_0$ given $\bZ_i$ is $f (\bt_0-\bbeta_0-\bbeta_1 \bZ_i)$,
where $f$ is the pdf (pmf) of $\mathbf{e}_i$; $i=1,\ldots,n$.
Without loss of generality, assume that $f(\mathbf{0})=M_0>0$. Since $\int f(\bt)d\bt=1$, for any $\epsilon_0>0$, there exists $r>0$ such that when some $\bt$ has length $\|\bt\| > r$, we have $0 \le f(\bt) < \epsilon_0$. Let $\epsilon_0=M_0/e^{\varepsilon}$.
Choose $\bz_i=\bbeta_1^{-1}(\bt_0-\bbeta_0)$ and $\bz_i'=\bbeta_1^{-1}(\bt_0-\bt-\bbeta_0)$. Then,
\ba
\frac{P(\fm(\bZ_i) = \bt_0|\bZ_i=\bz_i)}{P(\fm(\bZ_i) = \bt_0|\bZ_i=\bz_i')}=\frac{f(\mathbf{0})}{f(\mathbf{t})}>\frac{M_0}{\epsilon_0}=e^{\varepsilon}.
\ea
This completes the proof.

\subsubsection*{Proof of Theorem \ref{dp_multivar}}

This proof consists of four parts.

\noindent\textbf{Part 1: Univariate continuous distributions}

Note that $F(Z_i)$ follows a Uniform distribution $Unif(0,1)$, while $e_i$ follows an independent 
Laplace-distribution $Laplace(0,b)$. Then, $W_i=F(Z_i)+e_i$ follows the distribution $G$. Consequently, 
$G(W_i)$ follows a Uniform distribution $Unif(0,1)$ and thus $\tilde{Z}_i$ follows $F$. Then, $\tilde{Z}_1,\ldots,\tilde{Z}_n$ follow $F$ independently when we perform privation for 
each $Z_i$.

For $\varepsilon$-differential privacy, consider $\bz$ and $\bz'$ 
differ only in the $j$th element. Note that $e_j$ follows
the Laplace distribution $Laplace(0,1/\varepsilon)$. 
Then, the conditional pdf of $W_j|Z_j=z_j$ is 
$f_j(w_j|z_j)=\frac{\varepsilon}{2}e^{-\varepsilon|w_j-F(z_j)|}$ for any $w_j$,
implying that $\frac{f_j(w_j|z_j)}{f_j(w_j|z_j')} \le e^{\varepsilon|F(z_j)-F(z_j')|} \le e^{\varepsilon}$. Consequently, 
$\frac{f_\mathbf{W}(\mathbf{w}|\mathbf{z})}{f_{\mathbf{W}}(\mathbf{w}|\mathbf{z}')} = 
\frac{\prod_{i=1}^nf_i(w_i|z_i)}{\prod_{i=1}^nf_i(w_i|z_i')} = \frac{f_j(w_j|z_j)}{f_j(w_j|z_j')}\le e^{\varepsilon}$, where $f_\mathbf{W}(\mathbf{w}|\mathbf{z})$ is the conditional pdf
of $\mathbf{W}=\big(W_1,\ldots,W_n\big)$ given $\bZ=\mathbf{z}=\big(z_1,\ldots,z_n\big)$.  
As a consequence, for any measurable set $B$, 
\be \label{integral}
P(\mathbf{W} \in B|\mathbf{z}) = \int_B f_\mathbf{W}(\mathbf{w}|\mathbf{z}) d\mathbf{w} \le
\int_B e^{\varepsilon} f_{\mathbf{W}}(\mathbf{w}|\mathbf{z}')d\mathbf{w} = e^{\varepsilon}P(\mathbf{W} \in B|\mathbf{z}').
\ee
Then differential privacy of the proposed mechanism is guaranteed by Lemma 2.6 of \cite{wasserman2010statistical}. 

\noindent\textbf{Part 2: Univariate discrete distributions}

First, we prove that $\tilde{Z}_i$ follows $F$. Let $\tilde{V}_i= F_{V}^{-1}\circ G\big(F_{V}(V_i)+e_i\big)$. By Lemma \ref{lemma1}, $P(Z_i \le a_{k})=P(V_i \le a_{k})$;
$k =1,\ldots,s$, $i=1,\ldots,n$. By design, $V_i$ is a continuous variable. By Part 1 of this proof, $\tilde{V}_i$ follows the same distribution as $V_i$. Hence
$P(V_i \le a_{k})=P(\tilde{V}_i \le a_{k})$.
Therefore, we only need to show that $P(\tilde{V}_i \le a_{k})=P(\tilde{Z}_i \le a_{k})$, where $\tilde{Z}_i= \overline{L}(\tilde{V}_i) $. 
Since $\tilde{V}_i \le \overline{L}(\tilde{V}_i) $, two events 
$\{\overline{L} (\tilde{V}_i)  \le c\} \subseteq \{\tilde{V}_i \le c\} \mbox{ for any } c \in \mathbb{R}$. 
On the other hand, since the ceiling function maps $\tilde{V}_i$ to the lowest $a_{k}$ greater than or equal to $\tilde{V}_i$, we have
$\{\tilde{V}_i\le a_{k}\} \subseteq \{\overline{L}(\tilde{V}_i)  \le a_{k}\} \mbox{ for } k \in \{1,\ldots,s\}$.
This implies that $\{\tilde{V}_i\le a_{k}\} = \{\overline{L}(\tilde{V}_i)  \le a_{k}\}$ for $k \in \{1,\ldots,s\}$, and hence $P(\tilde{V}_i\le a_{k})=P(\tilde{Z}_i \le a_{k})$. 

Finally, we follow the same proof in Part 1 to establish $\varepsilon$-differential privacy for
$\tilde{\bV}=(\tilde{V}_1,\ldots,\tilde{V}_n)$ and then $\tilde{\bZ}=(\tilde{Z}_1,\ldots,\tilde{Z}_n)$. Details are omitted. 

\noindent\textbf{Part 3: Univariate mixed distributions}

From \eqref{ptb_mixed}, $P(Z_i \le x)=P(V_i \le x+k) =P(\tilde{V}_i \le x+k) = P(\tilde{Z}_i \le x)$ 
for $x=a_{k}$ and $x \in (a_{k},a_{k+1})$, $k=0,\pm 1,\ldots$. Meanwhile, we follow the same proof in Part 2 to establish $\varepsilon$-differential privacy for $\tilde{\bZ}$ and thus omit details.

\noindent\textbf{Part 4: Multivariate distributions}

Let $F$ and $\tilde{F}$ be the cdf's of the $p$-dimensional random variable $\bZ_i$ and $\tilde{\bZ}_i$, respectively. By the probability chain rule and Parts 1-3 of this proof, for any type of $F$, 
\ba
\tilde{F}(z_1,\ldots,z_p) &=& \tilde{F}(z_p|z_1,\ldots,z_{p-1}) \cdots \tilde{F}(z_2|z_1)
\tilde{F}(z_1)\\
&=&F(z_p|z_1,\ldots,z_{p-1}) \cdots F(z_2|z_1) F(z_1)  
= F(z_1,\ldots,z_p).
\ea 

To establish $\varepsilon$-differential privacy, consider $\bz=(\bz_1,\ldots,\bz_n)$ and 
$\bz'=(\bz_1',\ldots,\bz_n')$ differ only in the $j$th entry, and two realizations $\bz_j=(z_{j1},\ldots,z_{jp})$ and $\bz_j'=(z_{j1}',\ldots,z_{jp}')$ of
the $j$th random variable $\bZ_j$;
$j \in \{1,2,\ldots,n\}$. For simplicity, we first consider a measurable set of form $B=\prod_{j=1}^p B_j$
and we generalize the result to a general measurable set $B$ by approximating it by
a union of sets of the form. By the probability chain rule and \eqref{multi}, for $B=\prod_{j=1}^p B_j$,
\ba
P(\tilde{\bZ}_j \in B|\bz_j) = P(\tilde{Z}_{jp} \in B_p|\bz_j, \tilde{Z}_{j1},\ldots,\tilde{Z}_{j,p-1}) \cdots 
P(\tilde{Z}_{j2} \in B_2|\bz_j,\tilde{Z}_{j1}) P(\tilde{Z}_1 \in B_1|\bz_j).
\ea
	Applying the conclusion in Part 1, we obtain that 
	\ba
	P(\tilde{Z}_{jl} \in B_l|\bz_j, \tilde{Z}_{j1},\ldots,\tilde{Z}_{j,l-1})
	\le e^{\varepsilon/p} P(\tilde{Z}_{jl} \in B_l|\bz_j', \tilde{Z}_{j1},\ldots,\tilde{Z}_{j,l-1}).
	\ea
	As in \eqref{integral}, we integrate over the sample space of $z_{j1},\ldots,z_{j,l-1}$ and $z_{j1}',\ldots,z_{j,l-1}'$ with respect to the corresponding cdf's, and have
	\ba
	P(\tilde{Z}_{jl} \in B_l|z_{jl}, \tilde{Z}_{j1},\ldots,\tilde{Z}_{j,l-1})
	\le e^{\varepsilon/p} P(\tilde{Z}_{jl} \in B_l|z_{jl}', \tilde{Z}_{j1},\ldots,\tilde{Z}_{j,l-1}).
	\ea
	Therefore,
	\ba
	\frac{P(\tilde{\bZ}_j \in B|\bz_j)}{P(\tilde{\bZ}_j \in B|\bz_j')}
	&=& \frac{P(\tilde{Z}_{jp} \in B_p|z_{jp}, \tilde{Z}_{j1},\ldots,\tilde{Z}_{j,p-1})}{P(\tilde{Z}_{jp} \in B_p|z_{jp}', \tilde{Z}_{j1},\ldots,\tilde{Z}_{j,p-1})} \cdots
	\frac{P(\tilde{Z}_{j2} \in B_2|z_{j2},\tilde{Z}_{j1})}{P(\tilde{Z}_{j2} \in B_2|z_{j2}',\tilde{Z}_{j1})}
	\frac{P(\tilde{Z}_{j1} \in B_1|z_{j1})}{P(\tilde{Z}_{j1} \in B_1|z_{j1}')}\\
	&\le& e^{\varepsilon/p} \cdots 
	e^{\varepsilon/p}=e^{\varepsilon}.
	\ea



\subsubsection*{Proof of Proposition \ref{prop:multi_complexity}} 

For a univariate variable, converting $\bZ$ to $\bV$ and $\bZ^*$ to $\bV^*$ require $O(N)$ operations,
respectively. An application of \eqref{continualized_edf} requires sorting $\bV$ and $\bV^*$ jointly
to identify each $V_i$'s locations given the order statistics of $\bV^*$. This 
amounts to a complexity $O(N \log N)$ in the worst-case scenario \citep[e.g.,][]{cormen2009introduction}. Similarly, the computation of $\hat{C}^{-1}$ requires sorting observations for the empirical 
cumulative distributions, which amounts to a complexity $O(N \log N)$. Moreover, adding $e_i$ and applying $\hat{C}(\cdot)$, $G(\cdot)$, $\hat{C}^{-1}(\cdot)$, and $L(\cdot)$ require $O(N)$ operations. 
Altogether, the overall complexity for univariate privatization is $O(N \log N)$. 

For a $p$-dimensional variable, converting $\bZ$ to $\bV$ and $\bZ^*$ to $\bV^*$ requires $O(pN)$ 
operations. As in the univariate case, privatizing the first variable of $\bZ$ has a complexity $O(N \log N)$. 
For the second variable of $\bZ$, sorting the first variable of 
$(\bV,\bV^*)$ jointly has a complexity $O(N \log N)$. The actual calculation of $\hat{C}(V_{i2}|V_{i1})$, $i=1,\ldots,n$, in \eqref{multi_cond} has complexity $O(N)$ since each $V_{i2}$ is inserted in one of three intervals, as illustrated in Figure \ref{new_2d_continualization}(c), and basic operations follow. Then, all together, 
privatizing the second variable of $\bZ$
has complexity $O(N \log N)$.

For the third to the $p$th variable of $\bZ$, $l=3,\ldots,p$, if exist,
since the corresponding order statistics of $\bV^*$ are not identical almost surely, there exists at most one data point $\bV^*_{j_0l}$, $j_0 \in \{1,\ldots,m\}$, of which $V_{i1}$ is in the immediate interval on the left. 
Therefore, we only need to match $V_{i1}$ with the first variable of $\bV^*$, and the complexity is still $O(N \log N)$ almost surely. 
Then the privatization of all variables of $\bZ$ through applying \eqref{cond_priv} sequentially has complexity
$O(pN) + p \cdot O(N \log N) \sim O(p N \log N)$.
This completes the proof.

\subsubsection*{Proof of Theorem \ref{consistency_multivar}:} 

This proof consists of two parts.


\noindent\textbf{Part 1: Unknown univariate distributions}

Note that $\hat{C}^{-1}$ is not a function of data $\bZ$. Hence, by Lemma 2.6 of \cite{wasserman2010statistical}, we only need to investigate the $\epsilon$-differential privacy with respect to $\bW =  (\hat{C}(V_1) +e_1,\ldots,\hat{C}(V_n) +e_n)$. Meanwhile, $\hat{C} \in [0,1]$. Therefore, given the hold-out
sample $\bZ^*=\bz^*$, as in the proof of Theorem \ref{dp_multivar} part 1, 
we have
$P(\mathbf{W} \in B|\mathbf{z},\bz^*)  \le e^{\varepsilon}P(\mathbf{W} \in B|\mathbf{z}',\bz^*)$ for any measurable set $B$.
Note that $\bZ^*$ is independent of $\bZ$ and $\bZ'$. Then,
$$P(\mathbf{W} \in B|\mathbf{z}) = \int_{\bz^*}P(\mathbf{W} \in B|\mathbf{z},\bz^*)dP(\bz^*)  \le e^{\varepsilon} \int_{\bz^*} P(\mathbf{W} \in B|\mathbf{z}',\bz^*)dP(\bz^*) = P(\mathbf{W} \in B|\mathbf{z}'),$$
which satisfies the requirement of $\epsilon$-differential privacy.

Next, let $\tilde{F}$ be the distribution of the privatized sample $(\tilde{Z}_1,\ldots,\tilde{Z}_n)$, and let $C=F_V$ defined in \eqref{conti}. 
We first show that $\hat{C}$ is a consistent estimator of $C$ as $m \rightarrow \infty$, which leads to $\tilde{F}$ being a consistent estimator of $F$.

By the uniform convergence under the Kolmogorov-Smirnov distance, $\rho(\hat{F},F)=\sup_z |\hat{F}(z)-F(z)|=O(\frac{1}{\sqrt{m}})$ almost surely.
When $F$ is continuous,  by the construction of $\hat{C}$ in \eqref{continualized_edf}, $\rho(\hat{C},\hat{F}) \le m^{-1}$ almost surely. By the triangular inequality,
$\rho(\hat{C},C) \le \rho(\hat{C},\hat{F}) + \rho(\hat{F},F) =O(\frac{1}{\sqrt{m}})$ as $m \rightarrow \infty$, where
$F \equiv C$.
When $F$ is discrete, $\rho(\hat{F}_{\bV^*},C) =O(\frac{1}{\sqrt{m}})$ by Lemma \ref{lemma1},
where $\hat{F}_{\bV^*}$ is the empirical distribution function on $\bV^*$, that is, the continualized $\bZ^*$.
As in the continuous case, $\rho(\hat{C},\hat{F}_{\bV^*}) \le m^{-1}$, and 
$\rho(\hat{C},C) \le \rho(\hat{C},\hat{F}_{\bV^*}) + \rho(\hat{F}_{\bV^*},C) =O(\frac{1}{\sqrt{m}})$, which is the consistency of $\hat{C}$.

Finally, recall in \eqref{ptb_empirical2} that $\tilde{Z}_i = H\big(\hat{C}(V_i)+e_i\big)$ and $H=L \circ \hat{C}^{-1}\circ G$.
By the consistency of $\hat{C}$, the continuous mapping theorem, and the design of $L$, we conclude that $\tilde{Z}_i$ follows $F$ as $m \rightarrow \infty$,
which implies $\tilde{F}(z) \rightarrow F(z)$
for each $z \in \mathbb{R}$. By the Glivenko–Cantelli theorem (Theorem 19.1, \cite{vaart1998asymptotic}), 
$\rho(F,\tilde{F}) \rightarrow 0$ almost surely as $m \rightarrow \infty$.

\noindent\textbf{Part 2: Unknown multivariate distributions}

The result for $\varepsilon$-differential privacy follows by a direct application of Theorem \ref{dp_multivar} with $C$ replaced by $\hat{C}$ when applying \eqref{multi}.

Next, we establish consistency of $\tilde{F}$ for $F$. 
We first provide an explicit form of $\hat{C}$. Let $d_{jl}$ be the $j$th order statistic of the $l$th variable of $\bV^*$; $j=1,\ldots,m$, $l=1,\ldots,p$. 
And let $d_{0l}=\min_{j=1,\ldots,m} d_{jl} -1$.
Then we can define $\hat{C}$ as
\be \label{multivar_continualized_edf}
\hat{C}(\bx)=\sum_{k_1=0}^1\cdots \sum_{k_p=0}^1 \left(\prod_{l=1}^{p} \frac{|x_l-d_{j_{l}+k_l-1,l}|}{d_{j_ll}-d_{j_{l}-1,l}}\right) \hat{F}(d_{j_1-k_1,1},\ldots,d_{j_p-k_p,p}).
\ee
which generalizes \eqref{continualized_edf} to the high-dimensional case, and can be decomposed into a series of \eqref{multi_cond} based on the probability chain rule. 
Here $j_l \in \{1,\ldots,m\}$ is selected such that $x_l \in (d_{j_{l}-1,l}, d_{j_ll}]$; $l=1,\ldots,p$.
Note that 
$\hat{C}$ agrees with $\hat{F}$ on all observations in $\bZ^*$.

Suppose $F$ is continuous. By the construction of $\hat{C}$, 
$\rho(\hat{C},\hat{F}) \le m^{-1}.$ 
On the other hand, $\lim_{m \rightarrow \infty}\rho(\hat{F},F)=0$ almost surely is given by the Vapnik-Chervonenkis Theorem (Chapter 26 of \cite{shorack2009empirical}). 
By the triangular inequality, $\rho(\hat{C},F) \leq \rho(\hat{F},F)
+\rho(\hat{C}, \hat{F}) \rightarrow 0$ almost surely, 
implying the strong consistency of $\hat{C}$.
Suppose that the first $l$ variables of $\bZ^*$ are discrete. As in the continuous case, $\rho(\hat{C}, C) \rightarrow 0$.
As in the last paragraph of Part 1, 
$\tilde{F}(\bz) \rightarrow F(\bz)$ almost surely for each $\bz \in \mathbb{R}^p$. 

Next, we show that $\rho(\tilde{F},F) \rightarrow 0$ almost surely as $m \rightarrow \infty$. There exist $\bz_0, \bz_1, \ldots, \bz_K$ such that for each $k=1,\ldots,K$, we have $F(\bz_k)-F(\bz_{k-1})=1/K$, $z_{(k-1)l} \le z_{kl}$ for all $l=1,\ldots,p$, and $z_{(k-1)l'} < z_{kl'}$ for at least one $l' \in \{1,\ldots,p\}$.
Define $S_k = (-\infty, z_{k1}] \times \cdots \times (-\infty, z_{kp}].$
For any $\bz \in S_k\setminus S_{k-1}$, we have
$\tilde{F}(\bz) - F(\bz) \le  \tilde{F}(\bz_k) - F(\bz_{k-1}) = \tilde{F}(\bz_k) - F(\bz_{k}) +\frac{1}{K},$ and $\tilde{F}(\bz) - F(\bz) \ge \tilde{F}(\bz_{k-1}) - F(\bz_{k}) = \tilde{F}(\bz_{k-1}) - F(\bz_{k-1}) - \frac{1}{K}.$
This implies that, almost surely,
$\rho(\tilde{F},F) \le \max_{k=1}^K |\tilde{F}(\bz_k) - F(\bz_{k})| + \frac{1}{K}.$
Then $\tilde{F}(\bz) \rightarrow F(\bz)$ implies that for any integer $K$, $\max_{k} |\tilde{F}(\bz_k) - F(\bz_{k})| \rightarrow 0$.
This completes the proof.

{\subsubsection*{Proof of Lemma \ref{holdout}}
	By definition, $\bV_j$ is a continuous variable, and hence the (conditional) probability of any value is zero. This completes the proof.}

\subsubsection*{Proof of Lemma \ref{lemma1}}

Note that $V_i$ follows a mixture distribution. By the law of total probability, 
\ba
P(V_i \leq x)=\sum_{k=1}^{s} P(V_i \le x |Z_i=a_{k}) P(Z_i=a_{k})
=\sum_{k=1}^{s} \left(\int_{-\infty}^x \frac{\mathds{1}_{(a_{k-1}, a_{k}]}(v)}{a_{k}-a_{k-1}}dv \right) P(Z_i=a_{k})
\ea 
for any $x \in \mathbb{R}$, where $\mathds{1}_A(x)$ is an indicator function, with $\mathds{1}_A(x)=1$ 
if $x \in A$ and $\mathds{1}_A(x)=0$ otherwise. This completes the proof.

\subsubsection*{Proof of Lemma \ref{trans_LDP}:}
Note that $\tilde{Z}_i$ satisfies $\varepsilon$-local differential privacy. Then, for $i=1,\cdots,n$,
\ba
P(\zeta(\tilde{Z}_i) \in B_i |Z_i=z_i) &=& P(\tilde{Z}_i \in \zeta^{-1}(B_i) |Z_i=z_i) \\
&\le& e^{\varepsilon}P(\tilde{Z}_i \in \zeta^{-1}(B_i) |Z_i=z_i') = e^{\varepsilon}P(\zeta(\tilde{Z}_i) \in B_i |Z_i=z_i') 
\ea
for any measurable set $B_i$ and realizations $z_i, z_i'$; $i=1,\ldots,n$.
This completes the proof. 

\clearpage

\begin{figure}
	\centering
	\includegraphics[width=0.8\textwidth]{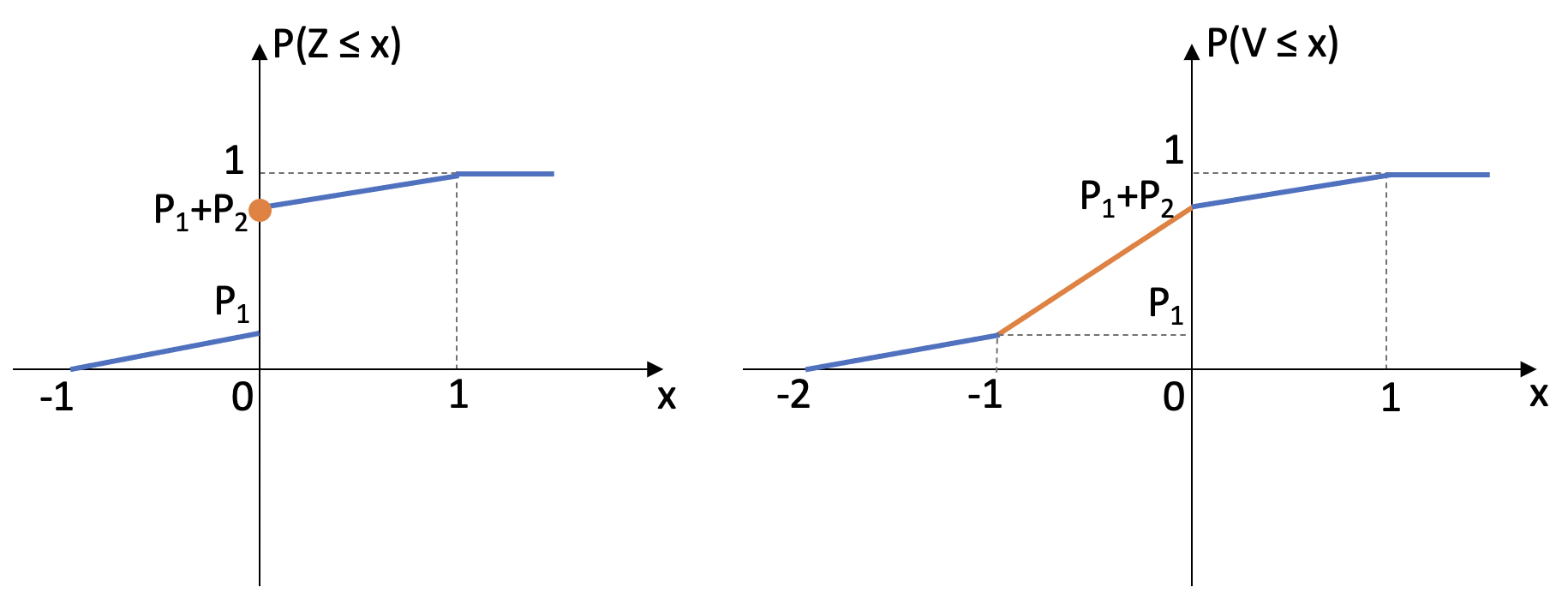}
	\caption{\textbf{Visualization of DIP for a mixed variable.} The left panel shows the cumulative distribution function of a mixed variable $Z$, which contains both continuous components (blue curves) and a discrete component (the orange jump at 0). Specifically, $Z=Unif(-1,0)$ with probability $P_1$, $Z=0$ with probability $P_2$, and $Z=Unif(0,1)$ with probability $1-P_1-P_2$. The right panel shows that the orange jump can be smoothed by inserting a linear piece on $(-1,0)$. The corresponding new variable $V$ is defined as $V=Z-1$ if $Z<0$, $V=Z-U$ if $Z=0$, and $V=Z$ if $Z>0$, where $U \sim Unif(0,1)$. This cumulative distribution function of $V$ is continuous and can be used for transforming back to $Z$.}
	\label{mixed_figure}
\end{figure}


\begin{figure}
	\centering
	\makebox[\linewidth][c]{
		\vspace{-.03in}
		\begin{subfigure}{.3\textwidth}\hspace{0.8in}
			\centering
			\includegraphics[width=\textwidth]{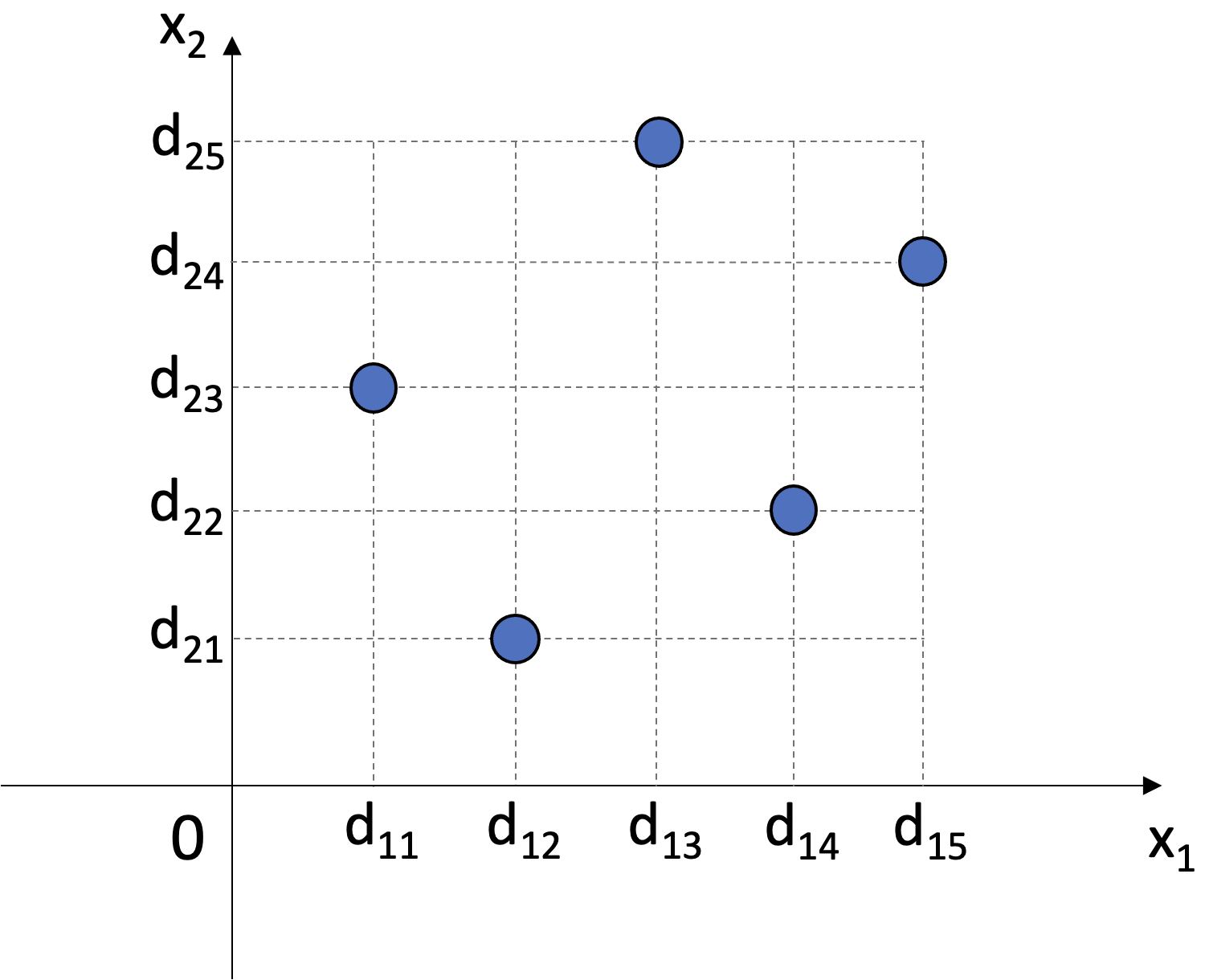}
			\caption{}
		\end{subfigure}
		\vspace{-.03in}
		\begin{subfigure}{.3\textwidth}
			\centering
			\includegraphics[width=\textwidth]{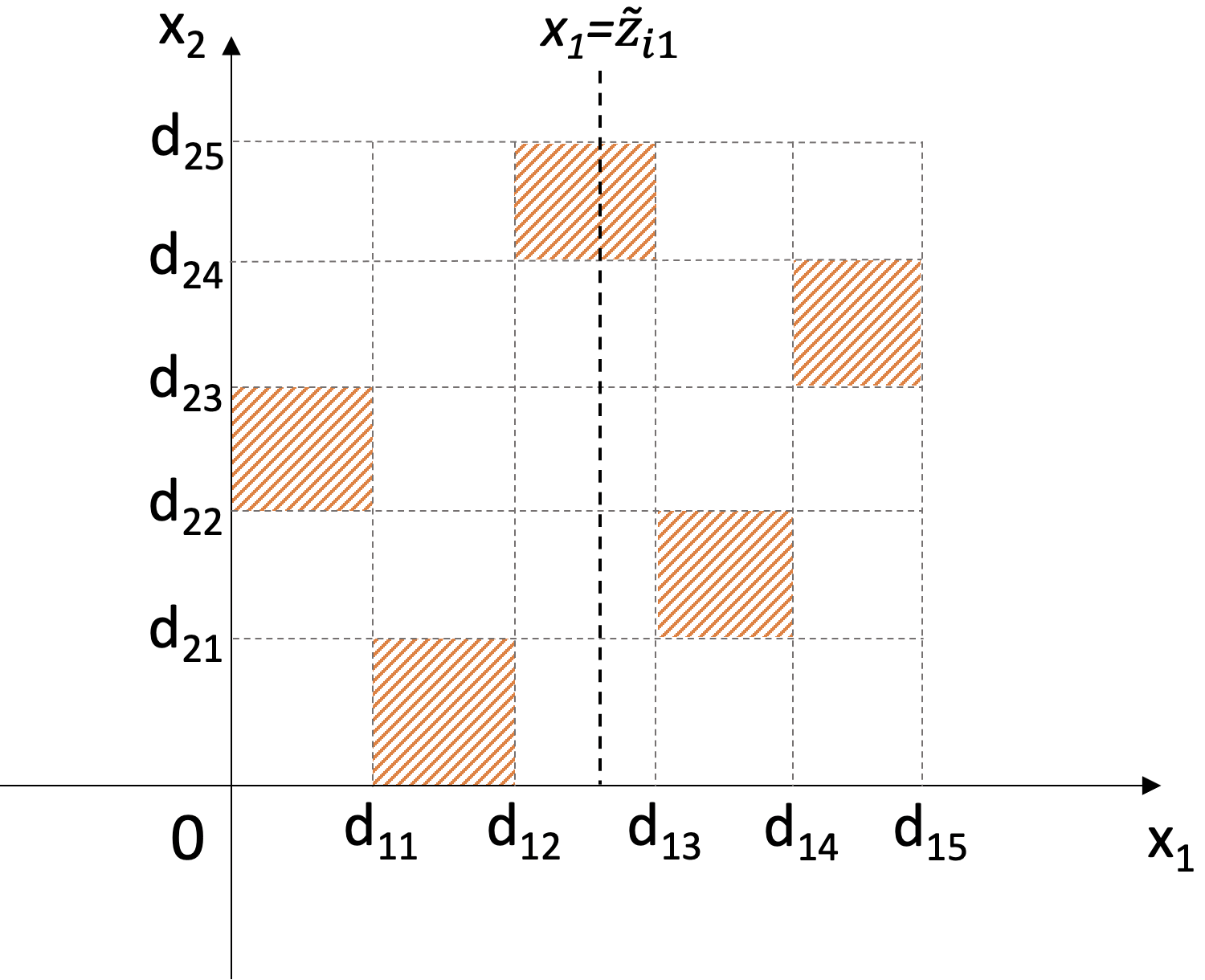}
			\caption{}
		\end{subfigure}
		\vspace{-.03in}
		\begin{subfigure}{.3\textwidth}
			\centering
			\includegraphics[width=\textwidth]{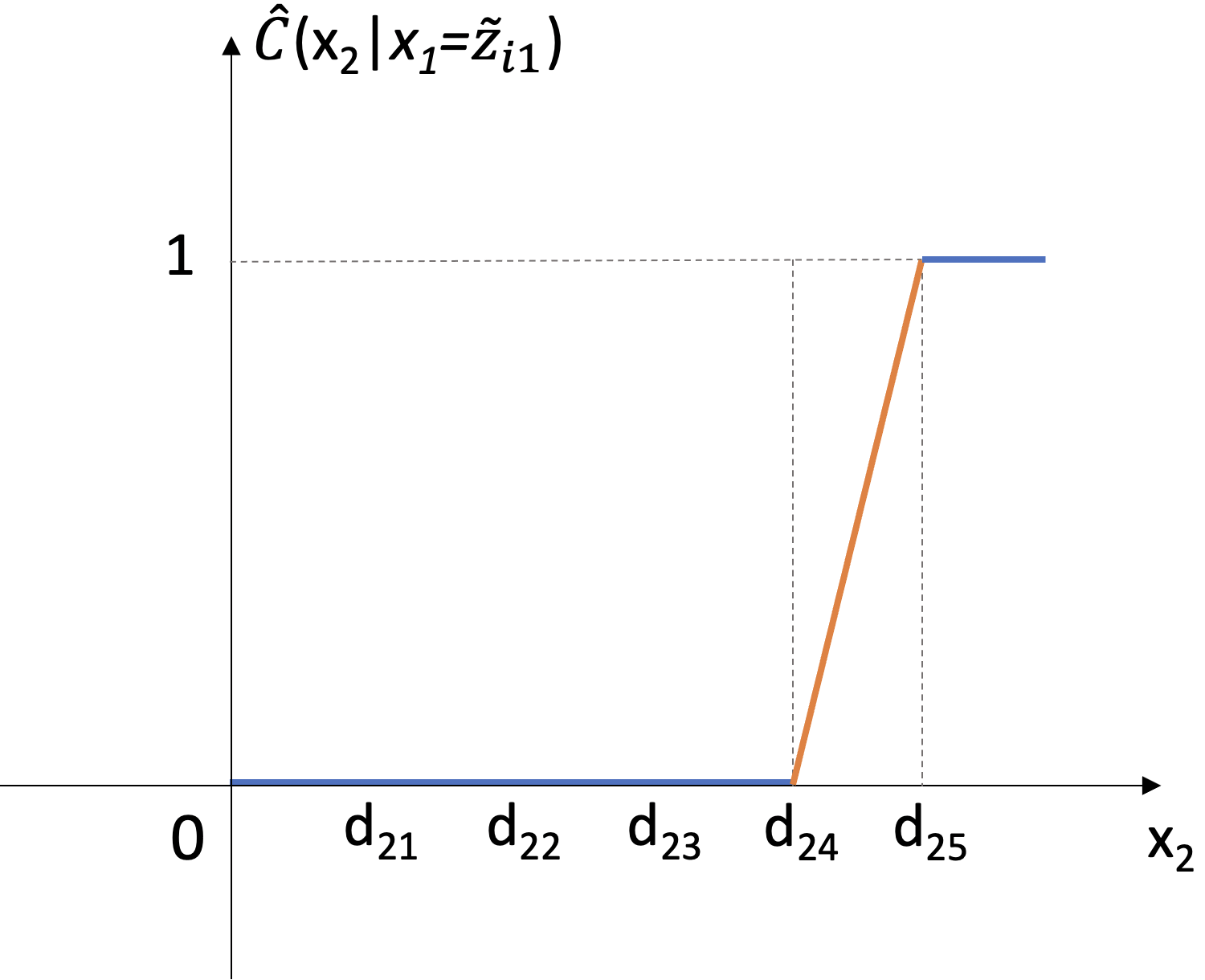}
			\caption{}
		\end{subfigure}
	}
	\caption{\textbf{Visualizations of a two-dimensional continualization for DIP.} (\textbf{a}) The probability mass function of an empirical two-dimensional variable $\hat{F}(x_1,x_2)$ (with jumps marked in blue dots). (\textbf{b}) The probability density function of its continualized version $\hat{C}(x_1,x_2)$ (with non-zero densities marked in orange shade), where for a privatized value $x_1=\tilde{z}_{i1}$, marked in the vertical dashed line, there is only one interval, $(d_{24},d_{25}]$, in which the density is non-zero. (\textbf{c}) The cumulative distribution function of $x_2$ given $x_1=\tilde{z}_{i1}$, which has a non-zero slope (i.e., probability) corresponding to the interval $(d_{24},d_{25}]$.}
	\label{new_2d_continualization}
\end{figure}

\clearpage
\begin{figure}[H]
	\centering
	\makebox[\linewidth][c]{
		\begin{subfigure}{.4\textwidth}
			\centering
			\includegraphics[width=\textwidth]{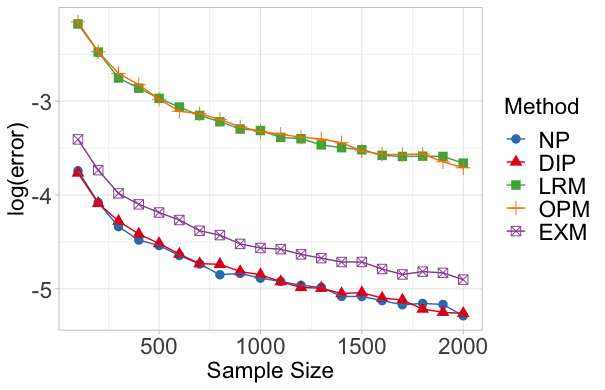}
			\caption{Bernoulli(0.1)}
		\end{subfigure}
		\vspace{-.03in}
		\begin{subfigure}{.4\textwidth}
			\centering
			\includegraphics[width=\textwidth]{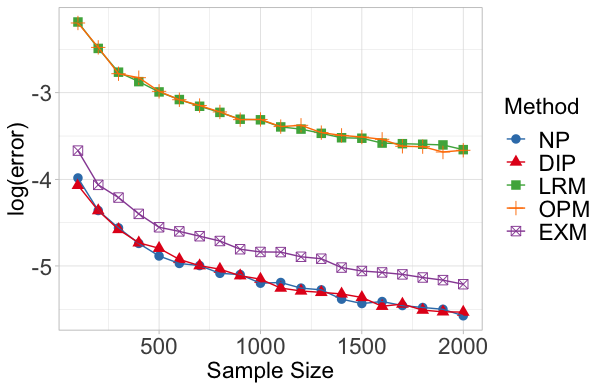}
			\caption{Binomial(5,0.5)}
		\end{subfigure}
	}
	
	\makebox[\linewidth][c]{
		\vspace{-.03in}
		\begin{subfigure}{.4\textwidth}\hspace{0.8in}
			\centering
			\includegraphics[width=\textwidth]{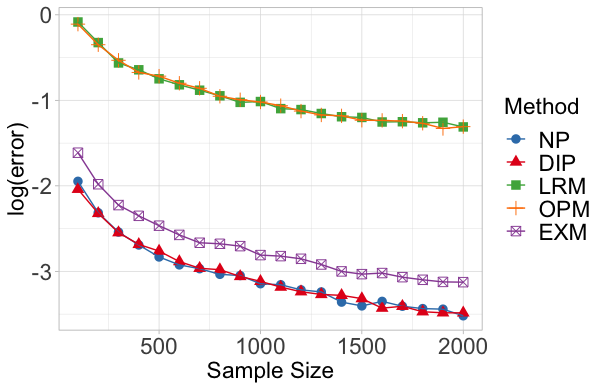}
			\caption{Poisson(3)}
		\end{subfigure}
		\vspace{-.03in}
		\begin{subfigure}{.4\textwidth}
			\centering
			\includegraphics[width=\textwidth]{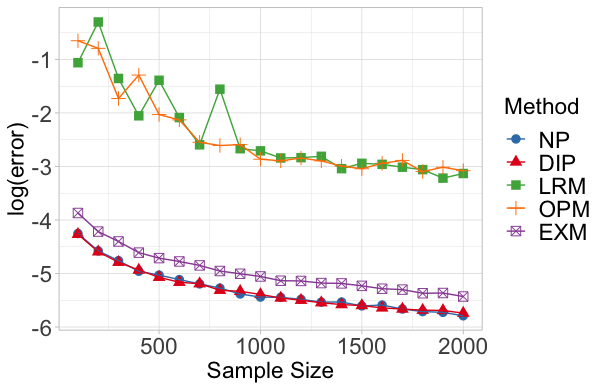}
			\caption{Geometric(0.2)}
		\end{subfigure}
	}
	\caption{
		\textbf{DIP performs as if the non-private data were used.} DIP is compared with the state-of-the-art privacy protection methods LRM, OPM, and EXM on simulated data, with data distributions provided in the sub-captions. The comparison concerns
		the mean estimation accuracy across different sample sizes under the same level of privacy protection strictness. A lower curve represents a smaller error. NP (in blue circles) represents the non-private benchmark. DIP (in red triangles) has curves almost align with those of NP, indicating nearly zero accuracy loss after privatization.} 
	\label{fig:ParaEst}
	\vspace{-1mm}
\end{figure}

%

\begin{table}
	\begin{footnotesize}
		\begin{center}
			\begin{tabular}{cccccc}
				\hline
				&& \multicolumn{4}{c}{$\varepsilon$}\\  \cmidrule{3-6}
				Distribution	&Method	&1 	&2 			&3  	& 4 \\
				\hline
				$Ber(0.1)$	&NP & 7.60 (5.67) & 7.47 (5.76) & 7.10 (5.59) & 7.93 (5.87) \\ 
				&DIP & 7.46 (5.69) & 7.99 (6.04) & 7.46 (5.67) & 7.50 (5.66) \\ 
				&LRM & 36.01 (28.78) & 19.56 (14.56) & 14.08 (11.01) & 11.94 (9.19) \\ 
				&OPM & 35.44 (27.24) & 20.19 (15.04) & 14.32 (11.03) & 12.39 (9.33) \\ 
				&EXM & 11.34 (8.16) & 10.43 (7.71) & 10.20 (7.76) & 11.05 (8.23) \\ 			
				\hline
				$Bin(5,0.5)$ &NP & 5.58 (4.20) & 5.68 (4.31) & 5.55 (4.20) & 5.56 (4.20) \\ 
				&DIP & 5.61 (4.29) & 5.89 (4.41) & 5.67 (4.27) & 5.86 (4.30) \\ 
				&LRM & 35.48 (28.46) & 19.00 (14.06) & 13.20 (10.08) & 10.70 (8.16) \\ 
				&OPM & 36.19 (27.11) & 18.91 (14.10) & 13.14 (10.03) & 10.51 (8.18) \\ 
				&EXM & 8.39 (6.26) & 7.83 (6.08) & 7.58 (5.93) & 7.90 (5.89) \\ 
				\hline
				$Pois(3)$ &NP & 43.23 (33.17) & 44.18 (33.14) & 43.05 (31.81) & 43.06 (32.45) \\
				&DIP & 43.99 (33.09) & 45.03 (34.32) & 44.36 (33.93) & 44.86 (33.23) \\
				&LRM & 354.48 (289.6) & 187.38 (141.77) & 127.02 (98.89) & 100.60 (77.04) \\
				&OPM & 352.31 (267.82) & 176.57 (133.99) & 123.39 (94.36) & 94.32 (75.18) \\
				&EXM & 64.63 (47.94) & 61.95 (46.27) & 60.70 (45.12) & 60.86 (45.77) \\
				\hline
				$Geom(0.2)$ &NP & 4.56 (3.36) & 4.58 (3.32) & 4.54 (3.44) & 4.44 (3.42) \\ 
				&DIP & 4.39 (3.43) & 4.35 (3.33) & 4.51 (3.32) & 4.47 (3.41) \\ 
				&LRM & 75.72 (300.3) & 25.52 (26.19) & 16.38 (13.81) & 13.03 (10.32) \\ 
				&OPM & 86.69 (695.74) & 25.16 (23.07) & 17.25 (14.75) & 12.66 (10.69) \\ 
				&EXM & 6.20 (4.79) & 6.47 (4.80) & 6.36 (4.75) & 6.30 (4.84) \\
				\hline
			\end{tabular}
			
		\end{center}
	\end{footnotesize}
	\caption{\textbf{DIP breaks the trade-off between differential privacy and statistical accuracy on discrete data protection.} Mean (and standard error) of the distance between the estimated and the true distributional parameters are presented. Different distributions of data (Bernoulli, Binomial, Poisson, Geometric) and privacy budgets (i.e., the privacy factor $\varepsilon$) are evaluated. DIP and the non-private (NP) benchmark are indistinguishable in most settings across different privacy budgets, indicating minimal accuracy loss while achieving privacy protection. The unit of values in the table is $10^{-3}$.}
	\label{sim2_para_est}
	
\end{table}

\begin{table}
	\begin{footnotesize}
		\begin{center}
			\begin{tabular}{cccccc}
				\hline
				&& \multicolumn{4}{c}{$\varepsilon$}\\  \cmidrule{3-6}
				Distribution	&Method	&1 	&2 			&3  	& 4 \\
				\hline
				$Unif(0,1)$	&NP	&27.09 (8.40) &28.01 (8.33) &26.91 (8.14) &27.13 (8.13) \\
				&DIP  	  &27.03 (8.01) 		&27.47 (8.15)		&27.39 (8.29)		&26.91 (7.85)\\
				&LRM 	 &326.54 (11.40) 		&224.93 (9.72)		&166.66 (9.49)		&129.55 (8.44)\\
				&OPM 	&510.39 (9.24) 		&508.39 (9.80)		&507.61 (9.33)		&506.56 (9.15)\\
				\hline
				$Beta(2,5)$ &NP	&27.70 (8.43) &26.85 (8.08) &27.13 (8.50) &27.19 (8.21) \\
				&DIP  	  &26.99 (7.94) 		&27.59 (8.06)		&27.51 (7.95)		&27.42 (8.12)\\
				&LRM 	  &384.62 (15.19) 		&304.66 (14.76)		&248.44 (13.62)		&205.59 (13.22)\\
				&OPM 	&510.14 (9.13) 		&508.38 (9.48)		&508.02 (9.72)		&507.38 (9.01)\\
				\hline
				$N(0,1)$ &NP	&27.16 (8.28) &27.08 (8.17) &26.73 (7.89) &27.86 (8.51) \\
				&DIP  	   	&27.10 (8.09) 		&27.24 (8.00)		&27.06 (8.02)		&27.08 (8.22)\\
				&LRM 	  	&368.71 (14.58) 		&285.12 (16.22)		&228.90 (17.14)		&188.22 (16.72)\\
				&OPM 	&507.85 (9.65) 		&504.92 (9.27)		&504.17 (9.90)		&503.34 (9.47)\\
				\hline
				$Exp(1)$  &NP	&27.37 (8.25) &27.31 (8.06) &27.51 (7.90) &27.65 (8.13) \\
				&DIP  	   	&27.62 (8.71) 		&26.97 (8.13)		&27.85 (8.46)		&27.92 (8.56)\\
				&LRM 	  	&439.61 (17.30) 		&392.35 (20.23)		&355.79 (22.60)		&323.98 (23.83)\\
				&OPM 	&510.54 (9.73) 		&508.62 (9.90)		&508.65 (9.79)		&508.29 (9.78)\\
				\hline
			\end{tabular}
			
		\end{center}
	\end{footnotesize}
	\caption{\textbf{DIP breaks the trade-off between differential privacy and statistical accuracy on continuous data protection.} Mean (and standard error) of Kolmogorov-Smirnov distance between the true distribution and the privatized 
		empirical distribution based on 1000 simulations. Four different distributions and privacy budgets are examined. DIP and the non-private (NP) benchmark are indistinguishable in most settings across different privacy budgets, indicating the minimal accuracy loss while achieving privacy protection. The unit of values is $10^{-3}$.}
	\label{sim1}
	
\end{table}

\begin{table}
	\begin{small}
		\begin{center}
			\begin{tabular}{cccccc}
				\hline
				\multicolumn{2}{c}{Graphical model}& \multicolumn{4}{c}{$\varepsilon$}\\  \cmidrule{3-6}
				Setting	&Method	&1 	&2 			&3  	& 4 \\
				\hline
				Chain Net	&NP & 0.01 (0.00) & 0.01 (0.00) & 0.01 (0.00) & 0.01 (0.00) \\
				$N=2000$ &{DIP (hold 15\%)} & 0.06 (0.02) & 0.06 (0.02) & 0.06 (0.02) & 0.06 (0.02) \\ 
				$p=5$ &{DIP (hold 25\%)} & 0.04 (0.01) & 0.04 (0.02) & 0.04 (0.02) & 0.04 (0.02) \\ 
				&{DIP (hold 35\%)} & 0.03 (0.01) & 0.03 (0.01) & 0.03 (0.01) & 0.03 (0.01) \\ 
				&NLRM & 37.08 (0.45) & 30.15 (0.44) & 26.08 (0.42) & 23.28 (0.45) \\
				&NOPM & 73.91 (0.43) & 67.07 (0.43) & 63.14 (0.41) & 60.54 (0.43) \\		
				\hline
				Chain Net	&NP & 0.06 (0.01) & 0.06 (0.01) & 0.06 (0.01) & 0.06 (0.01) \\
				$N=2000$ &{DIP (hold 15\%)} & 0.51 (0.07) & 0.51 (0.07) & 0.51 (0.08) & 0.51 (0.08) \\ 
				$p=15$ &{DIP (hold 25\%)} & 0.34 (0.05) & 0.34 (0.05) & 0.34 (0.05) & 0.34 (0.05) \\ 
				&{DIP (hold 35\%)} & 0.28 (0.04) & 0.28 (0.04) & 0.28 (0.04) & 0.28 (0.04) \\ 
				&NLRM & 145.84 (0.82) & 125.09 (0.79) & 112.92 (0.82) & 104.28 (0.78) \\
				&NOPM & 256.25 (0.79) & 235.54 (0.76) & 223.41 (0.80) & 214.84 (0.77) \\		
				\hline
				Exp Decay &NP & 0.01 (0.00) & 0.01 (0.00) & 0.01 (0.00) & 0.01 (0.00) \\
				$N=2000$ &{DIP (hold 15\%)} & 0.06 (0.02) & 0.06 (0.02) & 0.06 (0.02) & 0.06 (0.02) \\ 
				$p=5$ &{DIP (hold 25\%)} & 0.04 (0.02) & 0.04 (0.02) & 0.04 (0.02) & 0.04 (0.01) \\ 
				&{DIP (hold 35\%)} & 0.03 (0.01) & 0.03 (0.01) & 0.03 (0.01) & 0.03 (0.01) \\ 
				&NLRM & 34.39 (0.40) & 27.46 (0.40) & 23.41 (0.39) & 20.58 (0.40) \\
				&NOPM & 71.22 (0.38) & 64.38 (0.39) & 60.47 (0.38) & 57.84 (0.38) \\			
				\hline
				Exp Decay &NP & 0.06 (0.01) & 0.06 (0.01) & 0.06 (0.01) & 0.06 (0.01) \\
				$N=2000$ &{DIP (hold 15\%)} & 0.52 (0.08) & 0.51 (0.08) & 0.52 (0.08) & 0.51 (0.08) \\ 
				$p=15$ &{DIP (hold 25\%)} & 0.34 (0.05) & 0.34 (0.05) & 0.34 (0.05) & 0.34 (0.05) \\ 
				&{DIP (hold 35\%)} & 0.28 (0.04) & 0.28 (0.04) & 0.28 (0.04) & 0.28 (0.04) \\ 
				&NLRM & 136.11 (0.70) & 115.30 (0.69) & 103.16 (0.67) & 94.54 (0.67) \\
				&NOPM & 246.52 (0.67) & 225.75 (0.66) & 213.65 (0.64) & 205.10 (0.64) \\
				\hline
			\end{tabular}
			\vspace{1cm}
			
		\end{center}
	\end{small}
	\caption{\textbf{DIP breaks the trade-off between differential privacy and statistical accuracy on preserving network structures while providing differential privacy.} Entropy loss (standard error) between the true precision matrix and estimated precision matrix in a privatized Gaussian graphical model for chain and experiential decay networks is measured across different network structures, sizes, and privacy budgets. DIP's performance is the closest to the non-private benchmark, where a small estimation error is due to the unknown high-dimensional multivariate distribution.
	}
	\label{sim2:graphical}
	
\end{table}

\begin{table}
	\begin{small}
		
		\begin{center}
			\begin{tabular}{ccccc}
				\hline
				\hline
				& \multicolumn{4}{c}{$\varepsilon$}\\  \cmidrule{2-5}
				Method	&1	&2			&3	& 4 \\
				\hline
				\hline
				{DIP (hold 15\%)} & 0.50 (0.38) & 0.44 (0.34) & 0.40 (0.31) & 0.37 (0.28) \\ 
				{DIP (hold 25\%)} & 0.40 (0.31) & 0.36 (0.28) & 0.33 (0.25) & 0.30 (0.24) \\ 
				{DIP (hold 35\%)} & 0.36 (0.28) & 0.33 (0.26) & 0.31 (0.24) & 0.29 (0.22) \\ 
				LRM & 13.08 (9.95) & 6.65 (5.04) & 4.62 (3.43) & 3.37 (2.58) \\
				OPM & 4.69 (3.44) & 3.03 (2.31) & 2.34 (1.80) & 1.87 (1.39) \\
				\hline
				\hline
			\end{tabular}
			
		\end{center}
	\end{small}
	\caption{\textbf{DIP breaks the trade-off between differential privacy and statistical accuracy on the private average salary calculation using the University of California system data.} Average relative differences (i.e., $|\hat{\mu}-\mu|/\mu \times 100\%$) and standard error (in the parentheses) between the estimated and the true mean
		salary across different privacy budgets $\varepsilon$ are presented. DIP has minimal accuracy loss while achieving the same level of privacy protection.}
	\label{real1_mean}
	
\end{table}

\begin{table}
	\begin{small}
		\begin{center}
			\begin{tabular}{ccccc}
				\hline
				& \multicolumn{4}{c}{$\varepsilon$}\\  \cmidrule{2-5}
				Method	&1	&2			&3	& 4 \\
				\hline
				DIP (hold 15\%) & 4.80 (0.40) & 4.80 (0.40) & 4.80 (0.40) & 4.80 (0.40) \\ 
				DIP (hold 25\%) & 4.72 (0.31) & 4.73 (0.31) & 4.74 (0.32) & 4.72 (0.31) \\ 
				DIP (hold 35\%) & 4.69 (0.27) & 4.71 (0.28) & 4.69 (0.27) & 4.71 (0.28) \\ 
				NLRM & 31.07 (0.42) & 28.85 (0.40) & 26.81 (0.39) & 24.94 (0.37) \\ 
				NOPM & Inf & Inf & Inf & Inf \\
				\hline
			\end{tabular}
			
		\end{center}
	\end{small}
	\caption{\textbf{DIP breaks the trade-off between differential privacy and statistical accuracy on private logistic regression using bank marketing data.} Average Kullback-Leibler divergence (and standard error) between the estimated and the true probability in logistic regression are measured across different privacy budgets $\varepsilon$. DIP has minimal accuracy loss while achieving the same level of privacy protection. The unit of values is $10^{-2}$. Inf means infinity.}
	\label{real2_logistic}
	
\end{table}

\begin{table}
	\begin{small}
		\begin{center}
			\begin{tabular}{ccccc}
				\hline
				& \multicolumn{4}{c}{$\varepsilon$}\\  \cmidrule{2-5}
				Method	&1	&2			&3	& 4 \\
				\hline
				DIP (hold 15\%) & 1.03 (5.24 $\times 10^{-4}$) & 0.98 (8.16 $\times 10^{-4}$) & 0.94 (5.03 $\times 10^{-4}$) & 0.91 (4.67 $\times 10^{-4}$) \\ 
				DIP (hold 25\%) & 1.03 (4.96 $\times 10^{-4}$) & 0.98 (7.12 $\times 10^{-4}$) & 0.94 (5.27 $\times 10^{-4}$) & 0.92 (5.66 $\times 10^{-4}$) \\ 
				DIP (hold 35\%) & 1.04 (4.83 $\times 10^{-4}$) & 0.99 (9.16 $\times 10^{-4}$) & 0.95 (6.14 $\times 10^{-4}$) & 0.92 (6.24 $\times 10^{-4}$) \\ 
				LRM & 1.87 (1.03 $\times 10^{-3}$) & 1.31 (9.63 $\times 10^{-4}$) & 1.05 (9.77 $\times 10^{-4}$) & 0.93 (7.15 $\times 10^{-4}$) \\ 
				OPM & 2.61 (8.23 $\times 10^{-4}$) & 2.61 (8.07 $\times 10^{-4}$) & 2.61 (6.42 $\times 10^{-4}$) & 2.61 (7.74 $\times 10^{-4}$) \\
				EXM & 1.11 (5.52 $\times 10^{-4}$) & 1.08 (4.81 $\times 10^{-4}$) & 1.09 (4.86 $\times 10^{-4}$) & 1.09 (4.38 $\times 10^{-4}$) \\
				\hline
			\end{tabular}
			
		\end{center}
	\end{small}
	\caption{\textbf{DIP shows good performance on movie rating prediction using the MovieLens 25M data.} Average root mean square error (and standard error) between the predicted rating and the true rating in collaborative filtering on random test sets of the MovieLens 25M data across different privacy budgets $\varepsilon$. DIP has minimal prediction accuracy loss while achieving the same level of privacy protection.}
	\label{SVD}
	
\end{table}

\clearpage
\bibliographystyle{apalike}
\bibliography{diffpriv}

\end{document}